\documentclass[sigconf]{acmart}
\usepackage{enumitem}
\usepackage{balance}
\usepackage{algorithm}  
\usepackage{algpseudocode}
\usepackage{bm}
\usepackage{siunitx}
\usepackage{subcaption}
\usepackage{multirow}
\DeclareMathOperator*{\argmin}{arg\,min} 
\AtBeginDocument{%
  }

\copyrightyear{2024}
\acmYear{2024}
\setcopyright{rightsretained}
\acmConference[RecSys '24]{18th ACM Conference on Recommender
Systems}{October 14--18, 2024}{Bari, Italy}
\acmBooktitle{18th ACM Conference on Recommender Systems (RecSys '24),
October 14--18, 2024, Bari, Italy}\acmDOI{10.1145/3640457.3688148}
\acmISBN{979-8-4007-0505-2/24/10}

\begin{document}

\title{Accelerating the Surrogate Retraining for Poisoning Attacks against Recommender Systems}

\author{Yunfan Wu}
\affiliation{%
  \institution{Key Laboratory of AI Safety,}
  \country{Institute of Computing Technology, \\Chinese Academy of Sciences}
}
\affiliation{%
  \institution{University of Chinese Academy of}
  \country{Sciences, Beijing, China}
}
\email{wuyunfan19b@ict.ac.cn}

\author{Qi Cao}
\authornotemark[1]
\affiliation{%
  \institution{Key Laboratory of AI Safety, \\Institute of Computing Technology, \\Chinese Academy of Sciences}
  \country{Beijing, China}
}
\email{caoqi@ict.ac.cn}

\author{Shuchang Tao}
\affiliation{%
  \institution{Key Laboratory of AI Safety,}
  \country{Institute of Computing Technology, \\Chinese Academy of Sciences}
}
\affiliation{%
  \institution{University of Chinese Academy of}
  \country{Sciences, Beijing, China}
}
\email{taoshuchang18z@ict.ac.cn}

\author{Kaike Zhang}
\affiliation{%
  \institution{Key Laboratory of AI Safety,}
  \country{Institute of Computing Technology, \\Chinese Academy of Sciences}
}
\affiliation{%
  \institution{University of Chinese Academy of}
  \country{Sciences, Beijing, China}
}
\email{zhangkaike21s@ict.ac.cn}

\author{Fei Sun}
\affiliation{%
  \institution{Key Laboratory of AI Safety, \\Institute of Computing Technology, \\Chinese Academy of Sciences}
  \country{Beijing, China}
}
\email{sunfei@ict.ac.cn}

\author{Huawei Shen}
\affiliation{%
  \institution{Key Laboratory of AI Safety,}
  \country{Institute of Computing Technology, \\Chinese Academy of Sciences}
}
\affiliation{%
  \institution{University of Chinese Academy of}
  \country{Sciences, Beijing, China}
}
\email{shenhuawei@ict.ac.cn}

\renewcommand{\shortauthors}{Wu et al.}

\begin{abstract}
Recent studies have demonstrated the vulnerability of recommender systems to data poisoning attacks, where adversaries inject carefully crafted fake user interactions into the training data of recommenders to promote target items.
Current attack methods involve iteratively retraining a surrogate recommender on the poisoned data with the latest fake users to optimize the attack.
However, this repetitive retraining is highly time-consuming, hindering the efficient assessment and optimization of fake users.
To mitigate this computational bottleneck and develop a more effective attack in an affordable time, we analyze the retraining process and find that a change in the representation of one user/item will cause a cascading effect through the user-item interaction graph.
Under theoretical guidance, we introduce \emph{Gradient Passing} (GP),  a novel technique that explicitly passes gradients between interacted user-item pairs during backpropagation, thereby approximating the cascading effect and accelerating retraining.
With just a single update, GP can achieve effects comparable to multiple original training iterations.
Under the same number of retraining epochs, GP enables a closer approximation of the surrogate recommender to the victim. 
This more 
accurate approximation provides better guidance for optimizing fake users, ultimately leading to enhanced data poisoning attacks.
Extensive experiments on real-world datasets demonstrate the efficiency and effectiveness of our proposed GP.
\let\thefootnote\relax\footnotetext{*Corresponding author.}
\end{abstract}



\maketitle

\section{Introduction}
Recommender systems have become an essential component of modern online platforms, providing personalized recommendations that enhance user experience and engagement across various domains~\cite{goldberg1992using, covington2016deep, ying2018graph}.
Collaborative filtering (CF) is a widely adopted recommendation scenario, receiving extensive research attention~\cite{su2009survey}.
While the openness and collaborative nature of recommender systems offer convenience to users, they also render these systems vulnerable to adversarial attacks and manipulations~\cite{mobasher2007toward}, emphasizing the need for reliable and secure systems.

Adversaries conduct poisoning attacks by injecting crafted fake users into the training data of recommender systems~\cite{burke2005limited, li2016data, yang2017fake, zhang2021reverse}. 
In practice, fake accounts are registered for such manipulations, resulting in manipulated recommendations~\cite{zhang2023robust}. 
The business of selling fake YouTube views has been reported, highlighting the prevalence of adversary practices\footnote{https://www.nytimes.com/interactive/2018/08/11/technology/youtube-fake-view-sellers.html}.
Given the severe impacts, it is crucial to investigate poisoning attacks against recommender systems. 
Such research provides a foundation for developing robust defense and improving the trustworthiness of recommenders~\cite{zhang2023robust}.

\begin{figure}[t]
\centering
\begin{subfigure}{0.96\linewidth}
\includegraphics[width=\linewidth]{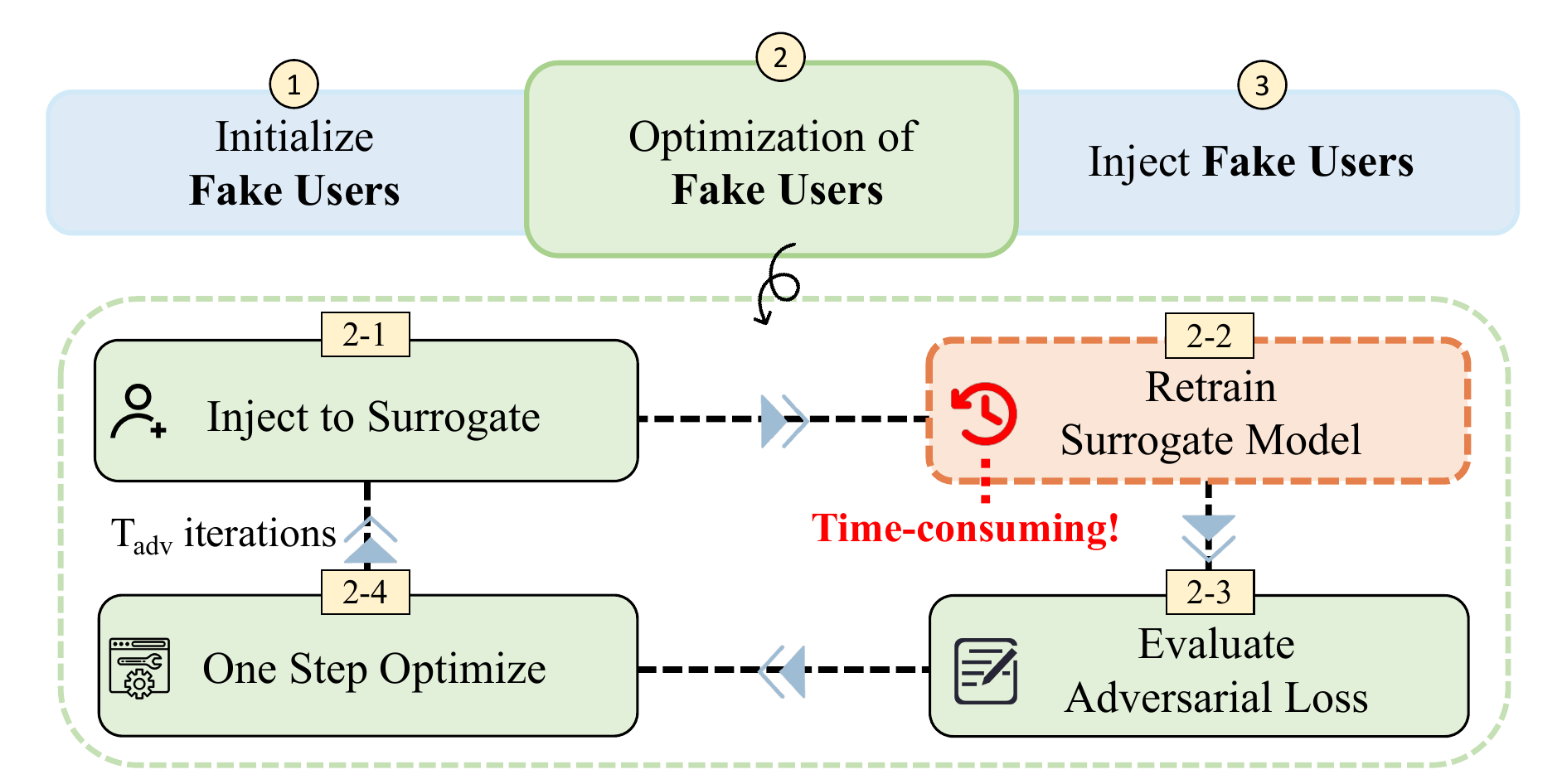}
\caption{General framework of optimization-based attacks.}
\label{fig:opt_framework}
\end{subfigure}
\begin{subfigure}{0.94\linewidth}
\includegraphics[width=\linewidth]{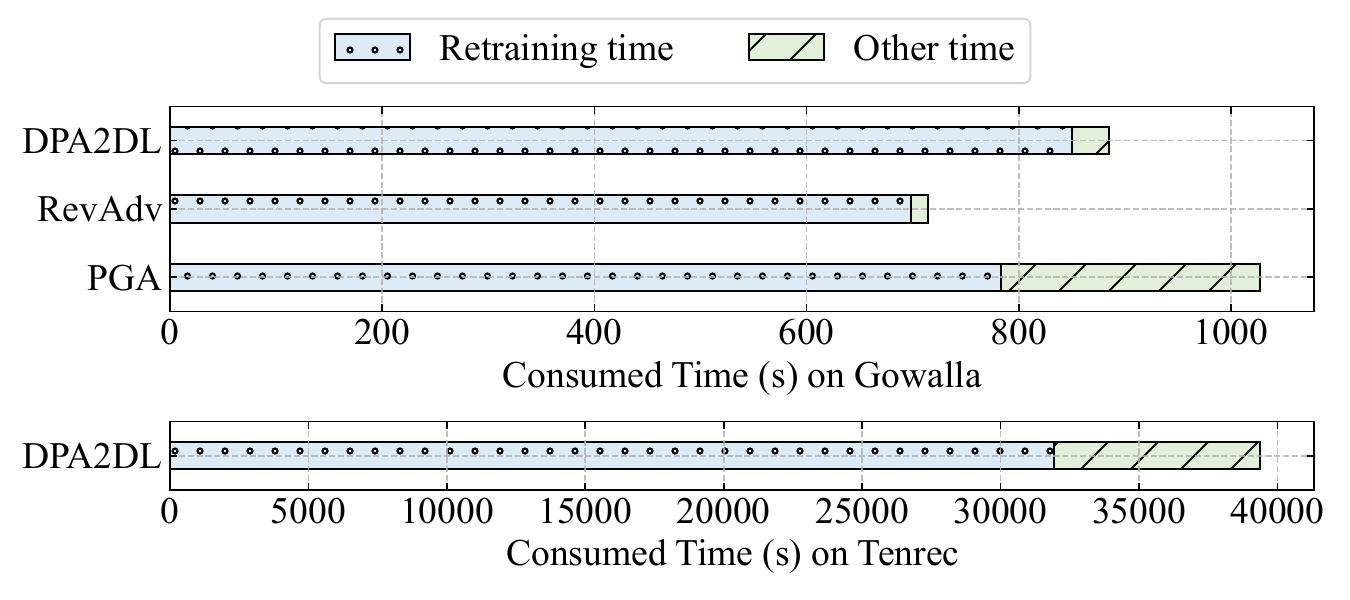}
\caption{Time consumed by existing attack methods.}
\label{fig:retrain_time}
\end{subfigure}
\caption{Retraining surrogate model is an important and time-consuming part of poisoning attacks.}
\end{figure}

Evolving from heuristic attack strategies, recent research has shifted its focus to optimization-based attacks~\cite{li2016data}.
These attacks iteratively optimize fake users by utilizing a surrogate recommender and an adversarial loss function. 
The surrogate recommender can evaluate fake users' attack effectiveness and guide the optimization to minimize the adversarial loss. 
After each update of fake users, the surrogate recommender has to be retrained on the poisoned data with the latest fake users (Figure~\ref{fig:opt_framework}). 
This repetitive surrogate retraining is the most time-consuming part of existing attack methods (Figure~\ref{fig:retrain_time}). 
It motivates us to investigate accelerating the surrogate retraining for more efficient and effective attacks.

Existing works mitigate this issue by restricting the retraining time~\cite{li2016data, tang2020revisiting, huang2021data}.
However, these approaches may reduce the overall attack effectiveness due to underdeveloped surrogate recommenders that behave differently from the victim.
Alternatively, some efforts have been made to avoid the retraining process by leveraging the influence function~\cite{zhang2020practical, wu2021triple, wu2023influence, koh2017understanding}.
Nevertheless, the influence function is originally designed to compute a data sample's impact on model training, assuming that the sample has been encountered during training.
So it is inaccurate to calculate the influence of a newly crafted adversarial sample without retraining.

In this study, we analyze the retraining process of CF models and find that the recommendation loss requires the representation similarity between interacted user-item pairs. 
As a result, updating a node's representation in current iteration triggers a cascading effect, affecting the representations of its connected nodes through the user-item interaction graph \textbf{in subsequent iterations}.

Inspired by this cascading dynamic, we propose \emph{Gradient Passing (GP)} to accelerate the surrogate retraining and enhance poisoning attacks.
During retraining, GP captures the changes of representations via gradients and explicitly passes them between interacted user-item pairs \textbf{within one iteration}, thereby approximating the cascading effect and accelerating model convergence.
Unlike the practice of message-passing in GNN-based recommenders during forward to improve expressiveness~\cite{wang2019neural}, we innovatively leverage gradients as messages in backward to enable faster retraining.

Both theoretical analysis and experiments demonstrate that one training iteration with GP can approximate the effects of multiple original iterations, significantly accelerating the surrogate retraining.
It allows for a closer approximation of the surrogate recommender to the victim, enhancing its accuracy in evaluating attack effectiveness and improving the optimization of fake users, which ultimately strengthens poisoning attacks.
Experiments on three real-world datasets verify that integrating GP into the state-of-the-art attack method can increase its average effectiveness by $29.57\%$, $18.02\%$, and $177.21\%$ while reducing the time cost by $43.27\%$, $40.54\%$, and $26.67\%$.

In this paper, we make the following contributions:
\begin{itemize}[leftmargin=*]
\item We introduce a novel method Gradient Passing (GP) based on both intuitive and theoretical analyses, accelerating the retraining process of surrogate recommenders.
\item We present the use of GP to enhance data poisoning attacks. It can be integrated into state-of-the-art attack methods and combined with other techniques.
\item Extensive experiments on three real-world datasets and six victim recommenders validate the efficiency and effectiveness of GP.
\end{itemize}

\section{Related Work}

\subsection{Recommender System}
Recommender systems have become ubiquitous in online applications in recent years, providing users with personalized suggestions~\cite{goldberg1992using, covington2016deep, ying2018graph}. 
Collaborative filtering (CF) is one of the most widely adopted recommendation tasks~\cite{su2009survey}. 
Its main objective is the top-$k$ recommendation, which aims to generate a personalized ranking list of $k$ items for each user. 
Early CF methods relied on similarity measures like Pearson correlation~\cite{resnick1994grouplens, sarwar2001item}, while more sophisticated latent factor models were later developed~\cite{hofmann2004latent, koren2009matrix, mnih2007probabilistic}. 
Recent advancements in deep learning have led to the development of neural CF models, such as autoencoders~\cite{sedhain2015autorec, liang2018variational}, convolutional neural networks~\cite{he2018outer}, and graph neural networks~\cite{wang2019neural, he2020lightgcn}.

This paper focuses on popular two-tower CF models, where two separate towers independently learn user and item representations to effectively capture complex user preferences and item characteristics~\cite{he2017neural, wang2019neural}. 
Once the latent representations are learned, user-item preference scores can be efficiently calculated using a similarity function like the dot product~\cite{ying2018graph}.

\subsection{Attack against Recommender System}
Early works on poisoning attacks focused on heuristic shilling attacks, such as random attacks~\cite{lam2004shilling}, bandwagon attacks~\cite{burke2005limited}, and others~\cite{mobasher2005effective, seminario2014attacking}. These attacks rely on the fundamental assumption of CF and generate fake users by heuristic rules. 
However, they are not specifically optimized for a recommendation model or an adversarial loss, leading to limited attack effectiveness. 
Recently, optimization-based attacks leveraged a selected surrogate model to obtain the attack feedback and minimize an adversarial loss.

Optimization-based attacks can be divided into two categories based on whether an attack model is utilized. 
\textbf{Model-based} attacks typically leverage Reinforcement Learning (RL)~\cite{zhang2020practical, song2020poisonrec, fan2021attacking, chen2022knowledge} or Generative Adversarial Networks (GANs) \cite{christakopoulou2019adversarial, lin2020attacking, wu2021triple, wang2023poisoning} to generate fake users. 
On the other hand, \textbf{model-free} approaches commonly employ adversarial gradients \cite{li2016data, fang2018poisoning, tang2020revisiting, fang2020influence, li2022revisiting, qian2023enhancing},  influence functions \cite{wu2023influence, huang2023single}, or other priors \cite{zhang2021reverse, zhang2021data, huang2021data, yue2021black, chen2023dark, wu2023attacking} to optimize fake users without learning an attack model.

The iterative retraining of surrogate recommenders remains the most time-consuming part of current poisoning attacks. 
In this paper, we investigate our proposed GP technique when integrated into RAPU-R~\cite{zhang2021data} and DPA2DL~\cite{huang2021data} for two reasons. 
First, training an attack model like RL or GAN is unstable and may be influenced by irrelevant factors. In contrast, RAPU-R and DPA2DL are model-free attacks that can better demonstrate the effectiveness of GP. 
Second, they not only achieve state-of-the-art attack performances but also scale well to large datasets. 
In contrast, other attacks that rely on high-order gradients or influence function can hardly be conducted when dealing with millions of users and items.

\subsection{Retraining of Recommender System}
In poisoning attacks, a common choice is restricting the time for surrogate retraining~\cite{li2016data, tang2020revisiting, huang2021data}.
However, it may reduce the overall attack effectiveness due to underdeveloped surrogate recommenders that behave differently from the victim.
Alternatively, some efforts have been made to avoid the repeated surrogate retraining by leveraging the influence function~\cite{zhang2020practical, wu2021triple, koh2017understanding}. 
Nevertheless, the influence function is originally designed to compute a data sample's impact on model training, assuming that the sample has been encountered during training.
So it is inaccurate to calculate the influence of a newly crafted adversarial sample without retraining.

The efficiency of retraining is also a concern in the field of incremental learning for recommender systems. 
In this field, researchers study the scenario where new feedback continually arrives. 
The core challenge lies in efficiently updating a previously trained recommender to maintain high performance on the latest data.
Incremental learning methods can be categorized into two main types: sample-based and model-based approaches~\cite{zhang2023survey}. 
Sample-based methods maintain a representative training sample set, circumventing the need to retrain on the entire large dataset and consequently reducing retraining time~\cite{diaz2012real}.
Model-based approaches employ a meta-learning model to directly update the parameters of recommendation models without retraining~\cite{zhang2020retrain}.
Drawing inspiration from incremental learning techniques, it has the potential to develop more efficient and effective poisoning attacks.

\section{Preliminaries}
This section introduces the fundamental concepts of recommender systems and formally defines data poisoning attacks against recommenders. 
We focus on item promotion attacks and use Hit Ratio as the measure of attack effectiveness. 
Table~\ref{tb:symbol} summarizes the important mathematical symbols used throughout the paper.

\begin{table}[t]
\centering
\caption{Summary of math symbols.}
\label{tb:symbol}
\renewcommand{\arraystretch}{1.}
\scalebox{0.9}{
\begin{tabular}{ll}
\hline
Symbol & Meaning \\
\hline
$\mathcal{U}, \mathcal{I}$ & Set of users $\vert \mathcal{U} \vert=n$, set of items $\vert \mathcal{I} \vert=m$ \\
$\mathsf{u}_i, \mathsf{i}_j$ & The user with index $i$, the item with index $j$ \\
$\bm{I}$ & User-item interaction matrix, $\bm{I} \in \{0,1\}^{n \times m}$\\
$\mathcal{I}_{\mathsf{u}} $ & Set of items interacted by user $\mathsf{u}$ \\
$\mathcal{U}_{\mathsf{i}} $ & Set of users who interacted with item $\mathsf{i}$ \\
$\bm{r}_\mathsf{u}, \bm{r}_\mathsf{i}$ & Representation vectors of user $\mathsf{u}$ and item $\mathsf{i}$, $\bm{r} \in \mathbb{R}^d$  \\
$\bm{R} $ & Representation matrix, $\bm{R} \in \mathbb{R}^{(n+m) \times d}$ \\
$\bm{g}_\mathsf{u}, \bm{g}_\mathsf{i}$ & Original gradient vectors of $\bm{r}_\mathsf{u}$ and $\bm{r}_\mathsf{i}$ \\
$\nabla_{\bm{R}} \mathcal{L}_\text{rec} $ & Original gradient matrix of $\bm{R}$ \\
$(\nabla_{\bm{R}} \mathcal{L}_\text{rec}) ^{\text{GP}}$ & Modified gradient matrix after GP \\
$\xi$ & Threshold controlling the scope of GP \\
$\overline{\bm{A}}^{\text{GP}}$ & Normalized GP matrix \\ 
$l$ & Number of GP layers \\
$\alpha$ & Coefficient controlling the weight of GP \\
\hline
\end{tabular}}
\end{table}

\subsection{Recommender System}
We formally define the components of a recommender system as follows. $\mathcal{U} = \{\mathsf{u}_1, \mathsf{u}_2, \dots, \mathsf{u}_n\}$ and $\mathcal{I} = \left\{ \mathsf{i}_1, \mathsf{i}_2, \dots, \mathsf{i}_m \right\}$ denote the sets of $n$ users and $m$ items, respectively. 
The user-item interaction matrix is represented as $\bm{I} \in \{0,1\}^{n \times m}$, where $\bm{I}_{i,j} = 1$ indicates user $\mathsf{u}_i$ has interacted with item $\mathsf{i}_j$, and 0 otherwise. 
We use $\mathcal{I}_{\mathsf{u}_i} = \{\mathsf{i}_j | \bm{I}_{i,j} = 1\}$ to denote the items interacted by user $\mathsf{u}_i$, and analogously $\mathcal{U}_{\mathsf{i}_j}$ for the users who have interacted with item $\mathsf{i}_j$.

Given the user set $\mathcal{U}$, item set $\mathcal{I}$, and interaction matrix $\bm{I}$, the recommendation model $\mathcal{M}$ learns a preference score $\mathcal{M}(\bm{\Theta}, \bm{I})=\bm{S} \in \mathbb{R}^{n \times m}$ for each user-item pair. The model parameters $\bm{\Theta}$ are optimized as:
\begin{equation}
\bm{\Theta}^{*} = \argmin_{\bm{\Theta}} \mathcal{L}_{\text{rec}}(\mathcal{M}(\bm{\Theta}, \bm{I}), \bm{I}),
\end{equation}
where $\mathcal{L}_{\text{rec}}$ is the recommendation loss.

This paper focuses on the top-$k$ recommendation task. 
For each user $\mathsf{u}_i \in \mathcal{U}$, the recommender identifies a set of items $\mathcal{T}_{\mathsf{u}_i} \subseteq (\mathcal{I} \setminus \mathcal{I}_{\mathsf{u}_i}) $ such that $|\mathcal{T}_{\mathsf{u}_i}| = k$, and for any item $\mathsf{i}_j \in \mathcal{T}_{\mathsf{u}_i}, \mathsf{i}_k \in \mathcal{I} \setminus (\mathcal{I}_{\mathsf{u}_i} \cup \mathcal{T}_{\mathsf{u}_i})$, we have $\bm{S}_{i, j} \ge \bm{S}_{i, k}$.

\subsection{Poisoning Attack against Recommenders}
We formalize the item promotion attack as follows. 
Let $\mathcal{U}^r$ and $\mathcal{U}^f$ denote the sets of real and fake users respectively, with $\vert \mathcal{U}^r \vert=n^r$ and $\vert \mathcal{U}^f \vert=n^f$. 
Given the interaction matrix of real users $\bm{I}^r \in \{0,1\}^{n^r \times m}$, a target item $\mathsf{i}_t$, and the victim recommendation model $\mathcal{M}_v$, data poisoning attacks aim to craft the interactions of $n^f$ fake users $\bm{I}^f \in \{0,1\}^{n^f \times m}$ under certain budget constraints. 
Then the generated fake interactions are injected into the training data of the victim recommender to promote the target item. 
Formally, the attack problem is defined as:
\begin{equation}
\begin{aligned}
&\max \limits_{\bm{I}^f}\ && \mathrm{HR}(\mathcal{M}_v(\bm{\Theta}^*, \bm{I}),\mathcal{U}^r,\mathsf{i}_t, k), \\
&\ \text{s.t.}\ && \bm{\Theta}^*= \argmin\limits_{\bm{\Theta}}\mathcal{L}_\text{rec}(\mathcal{M}_v(\bm{\Theta}, \bm{I}),\bm{I}), \\ 
&&& \bm{I} = \mathrm{concatenate}(\bm{I}^r, \bm{I}^f), \\
&&& \forall \mathsf{u} \in \mathcal{U}^f, \ \vert \mathcal{I}_\mathsf{u} \vert \le \tau.
\end{aligned}
\end{equation}
The objective is to maximize the Hit Ratio (HR) of the target item $\mathsf{i}_t$ among real users $\mathcal{U}^r$ in the victim recommender, which has been retrained using $\bm{I}$, while ensuring that the interactions of each fake user do not exceed a predefined budget $\tau$.
The poisoned interaction matrix $\bm{I} \in \{0,1\}^{(n^r + n^f) \times m}$ includes both real and fake users.

The HR evaluates the effectiveness of the data poisoning attack on a top-$k$ recommender system. It is defined as:
\begin{equation}
\mathrm{HR}(\mathcal{M}_v(\bm{\Theta}^*, \bm{I}), \mathcal{U}^r, \mathsf{i}_t, k) = \frac{1}{\vert \mathcal{U}^r \setminus \mathcal{U}_{\mathsf{i}_t} \vert} \sum_{\mathsf{u} \in \mathcal{U}^r \setminus \mathcal{U}_{\mathsf{i}_t} } \mathbb{I}(\mathsf{i}_t \in \mathcal{T}_{\mathsf{u}}),
\label{eq:hr}
\end{equation}
$\mathbb{I}(\cdot)$ is an indicator function, which is $1$ if the target item $\mathsf{i}_t$ is in $\mathcal{T}_{\mathsf{u}}$, and $0$ otherwise. 

\section{Methodology}
\begin{figure*}[t]
  \centering
  \includegraphics[width=0.94\linewidth]{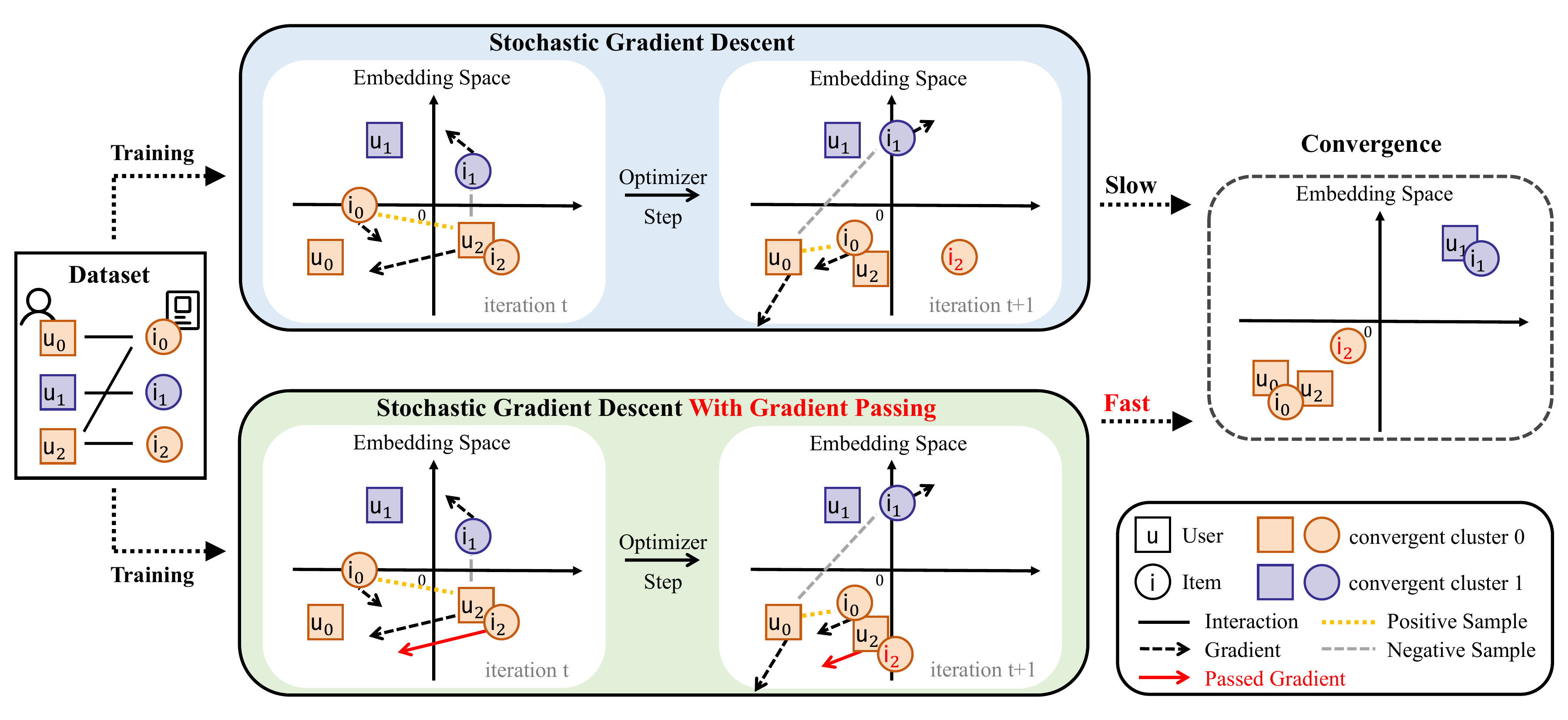}
  \caption{Representation optimization in a surrogate recommender over two iterations: comparing SGD alone (blue) to SGD with GP (green). GP accelerates retraining to the convergence state by passing gradients between interacted user-item pairs.}
  \label{fig:inspiration}
\end{figure*}


\subsection{Intuitive Discussion}
The two-tower architecture paired with dot product similarity is well-established in the field of CF~\cite{he2017neural, rendle2020neural}. 
These models maximize the similarity between interacted user-item pairs while minimizing non-interacted pairs. 
On this basis, we study how injected fake users influence the recommendation of real users.
When a fake user $\mathsf{u}$ is injected into the recommender system with representation $\bm{r}_{\mathsf{u}}$, its interacted item $\bm{r}_{\mathsf{i}}$ will be influenced to maximize their similarity.
This change in $\bm{r}_{\mathsf{i}}$ subsequently affects other users who have interacted with item $\mathsf{i}$, causing a cascading effect.

This cascading dynamic inspires GP, designed to accelerate the retraining of surrogate recommenders.
During training, gradients signify the direction and magnitude of changes required to minimize the recommendation loss.
Thus, gradients are the core signals in the cascading effect and we propose GP to explicitly pass them between interacted user-item pairs.
In this way, a single training iteration with GP could approximate the cascading effect in multiple original iterations and accelerate the convergence of recommenders.

We present an example with a toy dataset and two-dimensional vector representations for users and items, illustrated in Figure~\ref{fig:inspiration}.
Binary Cross Entropy (BCE) is used as the recommendation loss, with one positive and one negative sample per iteration.
At the $t^{\text{th}}$ iteration, GP passes gradients from $\bm{r}_{\mathsf{u}_2}$ to $\bm{r}_{\mathsf{i}_2}$ to preserve their high similarity.
The additional gradient information from GP guides $\bm{r}_{\mathsf{i}_2}$ to optimize towards its convergent cluster, accelerating the retraining. 
In contrast, with standard Stochastic Gradient Descent (SGD), $\bm{r}_{\mathsf{i}_2}$ not will be optimized until the positive pair $\mathsf{u}_2$ and $\mathsf{i}_2$ is sampled.

\subsection{Theoretical Analysis}
BCE loss is a commonly employed point-wise loss function \cite{he2017neural}, which formulates the recommendation task as a binary classification problem. 
The BCE loss with dot product similarity is:
\begin{equation}
\mathcal{L}_\text{rec} = \sum_{\mathsf{u}, \mathsf{i} \in \Omega} \mathrm{softplus} (-\bm{r}_{\mathsf{u}}^T \bm{r}_{\mathsf{i}}) + \beta \sum_{\mathsf{u}, \mathsf{i} \notin \Omega} \mathrm{softplus} (\bm{r}_{\mathsf{u}}^T \bm{r}_{\mathsf{i}}),
\end{equation}
with $\Omega = \left\{\left(\mathsf{u}_i, \mathsf{i}_j\right)| \bm{I}_{i,j} = 1\right\}$ representing the set of interacting pairs, and $\beta$ is the coefficient for negative samples. 
The softplus function is $\mathrm{softplus}(x) = \log(1 + e^x)$ or equivalently $-\log(\sigma(-x))$. 
Here $\bm{r}_{\mathsf{u}},  \bm{r}_{\mathsf{i}} \in \mathbb{R}^{d}$ are the vector representations of user $\mathsf{u}$ and item $\mathsf{i}$, and $\sigma$ represents the sigmoid function. 

\begin{lemma}
  \label{le:1}
  Let $\bm{R}=\text{vstack}(\bm{r}_{\mathsf{u}_1}, \cdots, \bm{r}_{\mathsf{u}_n}, \bm{r}_{\mathsf{i}_1}, \cdots, \bm{r}_{\mathsf{i}_m}) \in \mathbb{R}^{(n+m) \times d}$ denote the representation matrix for all users and items. 
  The gradient $\nabla_{\bm{R}} \mathcal{L}_\text{rec}$ can be derived through \textbf{message-passing} on $\bm{R}$. 
\end{lemma}

\begin{proof}
The gradient vector of $\mathcal{L}_\text{rec}$ with regard to the user representation $\bm{r}_{\mathsf{u}}$ is,
\begin{equation}
\bm{g}_{\mathsf{u}} = \nabla_{\bm{r}_{\mathsf{u}}} \mathcal{L}_\text{rec} = -\sum_{\mathsf{i} \in \mathcal{I}_{\mathsf{u}}} \sigma(-\bm{r}_{\mathsf{u}}^T \bm{r}_{\mathsf{i}}) \bm{r}_{\mathsf{i}} + \beta \sum_{\mathsf{i} \notin \mathcal{I}_{\mathsf{u}}} \sigma(\bm{r}_{\mathsf{u}}^T\bm{r}_{\mathsf{i}}) \bm{r}_{\mathsf{i}}.
\label{eq:gradient_ru}
\end{equation}
For brevity, $\bm{g}_{\mathsf{u}}$ and $\bm{g}_{\mathsf{i}}$ are used to denote $\nabla_{\bm{r}_{\mathsf{u}}} \mathcal{L}_\text{rec}$ and $\nabla_{\bm{r}_{\mathsf{i}}} \mathcal{L}_\text{rec}$.

Considering the components of $\bm{g}_{\mathsf{u}} \in \mathbb{R}^d$ from Equation (\ref{eq:gradient_ru}), $\bm{P}^{\text{grad}} \in \mathbb{R}^{n \times m}$ is constructed as:
\begin{equation}
\bm{P}^{\text{grad}}_{i, j}=
\begin{cases}
\sigma(-\bm{r}_{\mathsf{u}_i}^T \bm{r}_{\mathsf{i}_j}) &\mbox{if \ $\bm{I}_{i,j} = 1$}\\
- \beta \sigma(\bm{r}_{\mathsf{u}_i}^T \bm{r}_{\mathsf{i}_j}) &\mbox{otherwise}
\end{cases}.
\label{eq:p_grad}
\end{equation}
By comparing Equation (\ref{eq:gradient_ru}) with (\ref{eq:p_grad}), the relationship is established:
\begin{equation}
\text{vstack}(\bm{g}_{\mathsf{u}_1}, \cdots, \bm{g}_{\mathsf{u}_n}) = -\bm{P}^{\text{grad}} \text{vstack}(\bm{r}_{\mathsf{i}_1}, \bm{r}_{\mathsf{i}_2}, \cdots, \bm{r}_{\mathsf{i}_m}),
\end{equation}
where $\text{vstack}$ is a function that constructs an $n \times d$ matrix from $n$ vectors, each of size $d$.

Similarly, the following equation also holds. 

\begin{equation}
\text{vstack}(\bm{g}_{\mathsf{i}_1}, \cdots, \bm{g}_{\mathsf{i}_m}) = -(\bm{P}^{\text{grad}})^T \text{vstack}(\bm{r}_{\mathsf{u}_1}, \bm{r}_{\mathsf{u}_2}, \cdots, \bm{r}_{\mathsf{u}_n}),
\end{equation}

To calculate $\nabla_{\bm{R}} \mathcal{L}_\text{rec}=\text{vstack}(\bm{g}_{\mathsf{u}_1}, \cdots, \bm{g}_{\mathsf{u}_n}, \bm{g}_{\mathsf{i}_1}, \cdots, \bm{g}_{\mathsf{i}_m})$,  we construct $\bm{A}^{\text{grad}} \in \mathbb{R}^{(n+m) \times (n+m)}$ from $\bm{P}^{\text{grad}}$,
\begin{equation}
\bm{A}^{\text{grad}}  = 
\begin{pmatrix}
\ & \bm{P}^{\text{grad}} \\
(\bm{P}^{\text{grad}})^T &  \ \\ 
\end{pmatrix}.
\label{eq:extend}
\end{equation}

Finally, the gradient of $\mathcal{L}_\text{rec}$ with regard to $\bm{R}$ is,
\begin{equation}
\nabla_{\bm{R}} \mathcal{L}_\text{rec} = -\bm{A}^{\text{grad}} \bm{R} \in \mathbb{R}^{(n+m) \times d}.
\label{eq:gradient_r}
\end{equation} 
\end{proof}

Next we prove that \textbf{Gradient Passing} between user-item pairs could accelerate training for recommender systems.
\begin{proposition}
  \label{th:1}
  There exists a gradient passing matrix $\bm{A}^{\text{GP}} \in \mathbb{R}^{(n+m) \times (n+m)}$. When optimizing a recommender using BCE loss and SGD optimizer, a single iteration using $\bm{A}^{\text{GP}} \nabla_{\bm{R}} \mathcal{L}_{\text{rec}}$ can reach the state after two iterations with the original gradients $\nabla_{\bm{R}} \mathcal{L}_{\text{rec}}$.
\end{proposition}

\begin{proof}
Assuming that the SGD optimizer for the representation matrix $\bm{R}$ uses a learning rate $\alpha$, the update at iteration $t$ is:
\begin{equation}
\bm{R}_{t+1} = \bm{R}_t - \alpha \nabla_{\bm{R}} \mathcal{L}_\text{rec} (\bm{R}_t),
\end{equation}
where $\nabla_{\bm{R}} \mathcal{L}_\text{rec} (\bm{R}_t)$ is the gradient of $\mathcal{L}_\text{rec}$ w.r.t. $\bm{R}$ at iteration $t$. 

Using Equation (\ref{eq:gradient_r}), the update can be rewritten as:
\begin{equation}
\bm{R}_{t+1} = (\bm{1} + \alpha \bm{A}^{\text{grad}}_t)\bm{R}_t.
\end{equation}
Here, $\bm{1}$ represents the identity matrix.

Applying the update rule again for the subsequent step yields:
\begin{equation*}
\small
\begin{aligned}
\bm{R}_{t+2} & = (\bm{1} + \alpha\bm{A}^\text{grad}_{t+1})(\bm{1} + \alpha\bm{A^\text{grad}}_t)\bm{R}_{t} \\
& = \bm{R}_t + \alpha (\bm{1} + \alpha\bm{A}^\text{grad}_{t+1}) \bm{A}^\text{grad}_t \bm{R}_t + \alpha \bm{A}^\text{grad}_{t+1} \bm{R}_t \\
& = \bm{R}_t + \alpha \left[ (\bm{2} + \alpha\bm{A}^\text{grad}_{t+1}) \bm{A}^\text{grad}_t \bm{R}_t + (\bm{A}^\text{grad}_{t+1} -  \bm{A}^\text{grad}_{t}) \bm{R}_t \right] \\
& = \bm{R}_t - \alpha \left[\bm{2} + \alpha\bm{A}^\text{grad}_{t+1} + (\bm{A}^\text{grad}_{t+1} -  \bm{A}^\text{grad}_t) (\bm{A}^\text{grad}_t)^{-1} \right] \nabla_{\bm{R}} \mathcal{L}_\text{rec} (\bm{R}_t).
\end{aligned}
\end{equation*}

We define the gradient passing matrix as: 
\begin{equation}
\bm{A}^{\text{GP}} = \bm{2} + \alpha\bm{A}^\text{grad}_{t+1} + (\bm{A}^\text{grad}_{t+1} -  \bm{A}^\text{grad}_t) (\bm{A}^\text{grad}_t)^{-1}.
\label{eq:a_gp}
\end{equation}

The resulting update is: $\bm{R}_{t+2} = \bm{R}_{t} - \alpha \bm{A}^{\text{GP}} \nabla_{\bm{R}} \mathcal{L}_\text{rec} (\bm{R}_t)$.
Consequently, a single training iteration with passed gradients $\bm{A}^{\text{GP}} \nabla_{\bm{R}} \mathcal{L}_{\text{rec}}$ reaches the effect of two standard SGD iterations on $\bm{R}$.
\end{proof}

\subsection{Gradient Passing Strategy}
Ideally, the GP matrix $\bm{A}^{\text{GP}}$ would be defined as in Equation (\ref{eq:a_gp}). 
However, directly applying this formula faces several practical challenges. 
The matrix $\bm{A}^\text{grad}_{t+1}$ , required to compute $\bm{A}^{\text{GP}}$, is unknown at iteration $t$ due to its dependency on future state $\bm{R}_{t+1}$. 
Additionally, inverting $\bm{A}^\text{grad}_t$ poses computational difficulties. 
Moreover, the resultant $\bm{A}^{\text{GP}}$ is a dense matrix, whose use in GP would be computationally intensive with time complexity $\mathcal{O}(nmd)$.

While the exact application of $\bm{A}^{\text{GP}}$ is impractical, the equation provides valuable \textbf{theoretical guidance} for a feasible approach:
\begin{itemize}[leftmargin=*]
\item The matrix $\bm{A}^{\text{GP}}$ is composed of three terms: an identity matrix and two additional terms, carrying distinct weights, with the coefficients $2, \alpha$ and $1$ for three terms.
\item $\bm{A}^\text{grad}_{t+1}$ is block anti-diagonal, and $(\bm{A}^\text{grad}_{t+1} -  \bm{A}^\text{grad}_t) (\bm{A}^\text{grad}_t)^{-1}$ is block diagonal. The second term facilitates GP \emph{between user-item pairs}, while the third enables GP \emph{within user pairs and item pairs}.
\end{itemize}

Inspired by these insights, we first introduce a GP matrix $\bm{A}^{\text{GP-even}}$. Its \emph{even power} can pass gradients \emph{within user pairs and item pairs}, approximating the third term in Equation (\ref{eq:a_gp}). 

$\bm{A}^\text{grad}_{t+1} - \bm{A}^\text{grad}_t$ in the third term suggests, among interacted user-item pairs, the focus should be on those exhibiting a similarity reduction at iteration $t$.
Therefore, we introduce the condition term $\bm{r}_{\mathsf{u}}^T \bm{g}_{\mathsf{i}} + \bm{r}_{\mathsf{i}}^T \bm{g}_{\mathsf{u}}  > \xi_{\text{even}}$, where $\bm{g}_\mathsf{u}$ and $\bm{g}_\mathsf{i}$ represent the original gradients of $\bm{r}_\mathsf{u}$ and $\bm{r}_\mathsf{i}$. 
This condition is intrinsically interpreted as $\Delta(\bm{r}_{\mathsf{u}}^T \bm{r}_{\mathsf{i}})  < -\xi_{\text{even}}$. 
Specifically,
\begin{equation}
\begin{aligned}
- \Delta(\bm{r}_{\mathsf{u}}^T \bm{r}_{\mathsf{i}}) &  \approx -\bm{r}_{\mathsf{u}}^T \Delta(\bm{r}_{\mathsf{i}}) - \bm{r}_{\mathsf{i}}^T \Delta(\bm{r}_{\mathsf{u}}) \\
& \approx \bm{r}_{\mathsf{u}}^T \bm{g}_{\mathsf{i}} + \bm{r}_{\mathsf{i}}^T \bm{g}_{\mathsf{u}}.
\end{aligned}
\end{equation}
It is feasible because GP is performed during backpropagation when the original gradients have already been computed. 

Then, the subblock $\bm{P}^\text{GP-even}\in \mathbb{R}^{n \times m}$ of $\bm{A}^{\text{GP-even}}$ is defined:
\begin{equation}
\bm{P}^\text{GP-even}_{i,j}=
\begin{cases}
1 &\mbox{if \ $\bm{I}_{i,j} = 1$ and $\bm{r}_{\mathsf{u}}^T \bm{g}_{\mathsf{i}} + \bm{r}_{\mathsf{i}}^T \bm{g}_{\mathsf{u}} > \xi_{\text{even}}$}\\
0 &\mbox{otherwise}
\end{cases}.
\end{equation}

$\bm{A}^{\text{GP-even}}$ is extended from $\bm{P}^\text{GP-even}$ as Equation (\ref{eq:extend}). 
Next, we normalize it following GCN~\cite{kipf2016semi}:
\begin{equation}
\overline{\bm{A}}^\text{GP-even} = \bm{D}^{-1/2}\bm{A}^\text{GP-even}\bm{D}^{-1/2},
\end{equation}
where $\bm{D}=\text{diag}(\vert \mathcal{I}_{\mathsf{u}_1} \vert, \cdots, \vert \mathcal{I}_{\mathsf{u}_n} \vert, \vert \mathcal{U}_{\mathsf{i}_1} \vert, \cdots,  \vert \mathcal{U}_{\mathsf{i}_m} \vert) $ is a diagonal degree matrix, representing the number of interactions.

We apply $\bm{2l}$ message passing layers to the original gradient $\nabla_{\bm{R}} \mathcal{L}_\text{rec}$, which are defined as:
\begin{equation}
\begin{aligned}
(\nabla_{\bm{R}} \mathcal{L}_\text{rec})^\text{GP-even}_0 &= \nabla_{\bm{R}} \mathcal{L}_\text{rec}, \\
(\nabla_{\bm{R}} \mathcal{L}_\text{rec})^\text{GP-even}_{i+1} &= \overline{\bm{A}}^\text{GP-even} (\nabla_{\bm{R}} \mathcal{L}_\text{rec})^\text{GP-even}_i.
\end{aligned}
\end{equation}
Remember that, $\bm{A}^{\text{GP-even}}$ is to pass gradients \emph{within user pairs and item pairs}. So we only collect $l$ even index terms and obtain 

\begin{equation}
\bm{G}^\text{GP-even} = \sum_{i=1}^l (\nabla_{\bm{R}} \mathcal{L}_\text{rec})^\text{GP-even}_{2i}.
\end{equation}

To pass gradients \emph{between user-item pairs}, we construct $\bm{A}^{\text{GP-odd}}$ similarly and collect its odd terms $\bm{G}^\text{GP-odd} = \sum_{i=1}^l (\nabla_{\bm{R}} \mathcal{L}_\text{rec})^\text{GP-odd}_{2i - 1}$. 
The only distinction between $\bm{A}^{\text{GP-odd}}$ and $\bm{A}^{\text{GP-even}}$ lies in their different thresholds, $\xi_{\text{odd}}$ and $\xi_{\text{even}}$. 

Finally, we assign the odd and even terms with different weights $\alpha_{\text{odd}}, \alpha_{\text{even}}$ and modify the gradients of  $\bm{R}$ for the optimizer to perform gradient descent.
\begin{equation}
(\nabla_{\bm{R}} \mathcal{L}_\text{rec})^\text{GP} = \nabla_{\bm{R}} \mathcal{L}_\text{rec} + \alpha^{\text{odd}} \bm{G}^\text{GP-odd} + \alpha^{\text{even}} \bm{G}^\text{GP-even}.
\end{equation}

In our proposed GP strategy, four hyperparameters are introduced: $\xi_{\text{odd}}$, $\xi_{\text{even}}$, $\alpha_{\text{odd}}$, and $\alpha_{\text{even}}$. 
$\xi$ is designed to control the gradient passing scope and $\alpha$ determines the weight.

GP can be incorporated to enhance existing poisoning attacks by enabling a closer approximation of the surrogate recommender to the victim. 
The specific procedure of a state-of-the-art attack DPA2DL~\cite{huang2021data} enhanced by GP is shown in Algorithm \ref{alg:attack}. 

\begin{algorithm}[t]
  \caption{Enhanced DPA2DL Attack via Gradient Passing}  
  \label{alg:attack}  
  \begin{algorithmic}[1]
    \small
    \Require  
      Real user interactions $\bm{I}^r \in \{0,1\}^{n^r \times m}$; target item $\mathsf{i}_t$; surrogate model $\mathcal{M}_s$; recommendation loss $\mathcal{L}_\text{rec}$; adversary loss $\mathcal{L}_\text{adv}$; attack budgets $n^f$, $\tau$; GP hyperparameters $l$, $\xi$, $\alpha$; DPA2DL hyperparameters.
    \Ensure  
      Fake user interactions $\bm{I}^f \in \{0,1\}^{n^f \times m}$.
    \State \textbf{Initialize} $\bm{I}^f$ as a zero matrix of size $n^f \times m$.
    \For{$i = 0$ \textbf{to} $n^f-1$}
      \State  \textcolor{gray}{// The $i^{\text{th}}$ row $\bm{I}^f_i$ is the interaction vector for the $i^{\text{th}}$ fake user $\mathsf{u}^f_i$}.
      \State Add an interaction between $\mathsf{u}^f_i$ and the target item $\mathsf{i}_t$ to $\bm{I}^f_i$.
      \State Initialize parameters $\bm{\Theta}_s$ for $\mathcal{M}_s$ with user size $n^r+i+1$.
      \State Train $\mathcal{M}_s$ on poisoned dataset $[\bm{I}^r; \bm{I}^f]$ using $\mathcal{L}_\text{rec}$, apply GP.
      \State Train $\mathcal{M}_s$ with both $\mathcal{L}_\text{rec}$ and $\mathcal{L}_\text{adv}$, apply GP on $\mathcal{L}_\text{rec}$.
      \State Predict preference scores $\bm{S}^f_i$ for $\mathsf{u}^f_i$ using $\mathcal{M}_s$.
      \State Update $\bm{I}^f_i$ based on $\bm{S}^f_i$ with at most $\tau$ interactions.
    \EndFor
    \State \textbf{return} $\bm{I}^f$.
  \end{algorithmic}  
\end{algorithm}

\textbf{Complexity Analysis.}
The time complexity of GP is $\mathcal{O}(\Vert \bm{I} \Vert_0 ld)$. 
Here, $\Vert \bm{I} \Vert_0$ is the number of interactions. $l$ signifies the number of GP layers, and $d$ is the hidden size of user/item representations. 

\section{Experiments}
To thoroughly evaluate the effectiveness of GP in accelerating surrogate retraining and enhancing poisoning attacks, we conduct extensive experiments to analyze the following questions:
\begin{itemize}[leftmargin=*]
\item \textbf{Q1:} How does GP enhance the efficiency and effectiveness of state-of-the-art poisoning attacks? 
\item \textbf{Q2:} Does GP maintain its effectiveness when pre-training a surrogate model and combined with other techniques?
\item \textbf{Q3:} Do real-world examples support our motivations and how do hyper-parameters influence the effectiveness of GP?
\end{itemize}

\subsection{Experimental Settings}

\subsubsection{Datasets}
We conduct experiments on three publicly available benchmark datasets: \textbf{Gowalla} \footnote{\url{https://snap.stanford.edu/data/loc-Gowalla.html}} \cite{cho2011friendship}, \textbf{Yelp} \footnote{\url{https://www.yelp.com/dataset}}, and \textbf{Tenrec} \footnote{\url{https://static.qblv.qq.com/qblv/h5/algo-frontend/tenrec_dataset.html}} \cite{yuan2022tenrec}, which represent diverse domains with items corresponding to geographical locations, local businesses, and news articles. To ensure data quality and align with previous work \cite{tang2020revisiting}, we pre-process the datasets by filtering out users and items with fewer than $15$ interactions. For each remaining user, its interactions are chronologically split into training ($80\%$) and validation ($20\%$) sets for training and hyper-parameter tuning of recommenders. For Yelp dataset, we consider ratings above $3$ as interactions. For Tenrec dataset, we treat clicks as interactions. Key statistics of the processed datasets are summarized in Table~\ref{tb:dataset}.

\begin{table}[t]
  \caption{Statistics of the datasets.}
  \label{tb:dataset}
  \centering
   \renewcommand{\arraystretch}{0.95}
  \begin{tabular}{lS[table-format=7.0]S[table-format=5.0]S[table-format=8.0]S[table-format=0.5]}
    \toprule
    Dataset & {\#Users} & {\#Items} & {\#Interactions} & {Density\%} \\
    \midrule
    Gowalla & 13149 & 14009 & 535650 & 0.29079 \\
    Yelp & 35528 & 24573 & 1268345 & 0.14528\\
    Tenrec & 1195207 & 97761 & 40806690 & 0.03492\\
    \bottomrule
\end{tabular}
\end{table}

\begin{table*}
  \caption{Data poisoning attack evaluated by $\text{Recall}@50 (\%)$, the most effective attack is highlighted in bold.}
  \label{tb:attack}
  \scalebox{0.9}{
  \renewcommand{\arraystretch}{0.4}
  \begin{tabular}{ccccccccc}
    \toprule
   Dataset & Attacker & {MF-BPR} & {MF-APR} & {LightGCN} & {MultiVAE} & {NeuMF} & {MF-BCE} & {Average\textsuperscript{1}} \\
   \midrule
   \multirow{11}{*}{Gowalla} & None & $0.423 \pm 0.344$ & $0.427 \pm 0.460$ & $0.409 \pm 0.353$ & $0.332 \pm 0.295$ & $0.388 \pm 0.290$ & $0.304 \pm 0.326$ & $0.380 \pm 0.339$\\
   \cmidrule{2-9}
   & Random & $0.280 \pm 0.204$ & $0.178 \pm 0.237$ & $1.131 \pm 0.217$ & $0.332 \pm 0.360$ & $0.462 \pm 0.224$ & $0.265 \pm 0.284$ & $0.441 \pm 0.247$\\
   & Bandwagon & $0.386 \pm 0.265$ & $0.302 \pm 0.316$ & $\mathbf{1.748 \pm 0.162}$ & $0.385 \pm 0.404$ & $0.469 \pm 0.275$ & $0.233 \pm 0.349$ & $0.587 \pm 0.284$ \\
   \cmidrule{2-9}
   & PGA & $0.528 \pm 0.317$ & $0.409 \pm 0.186$ & $0.777 \pm 0.199$ & $0.544 \pm 0.499$ & $0.520 \pm 0.231$ & $0.373 \pm 0.323$ & $0.525 \pm 0.285$\\
   & RevAdv & $0.521 \pm 0.345$ & $0.423 \pm 0.260$ & $0.683 \pm 0.270$ & $0.472 \pm 0.385$ & $0.512 \pm 0.277$ & $0.369 \pm 0.296$ & $0.496 \pm 0.302$\\
   \cmidrule{2-9}
   & RAPU-R & $0.614 \pm 0.325$ & $0.588 \pm 0.258$ & $0.858 \pm 0.287$ & $0.498 \pm 0.420$ & $0.617 \pm 0.353$ & $0.587 \pm 0.546$ & $0.627 \pm 0.361$ \\
   & RAPU-R$\times2$ & $0.867 \pm 0.472$ & $0.909 \pm 0.403$ & $0.842 \pm 0.384$ & $0.755 \pm 0.549$ & $0.766 \pm 0.458$ & $0.943 \pm 0.583$ & $0.847 \pm 0.448$ \\
   & RAPU-R+GP & $0.945 \pm 0.513$ & $1.086 \pm 0.561$ & $0.656 \pm 0.285$ & $0.987 \pm 0.508$ & $0.854 \pm 0.482$ & $1.086 \pm 0.701$ & $0.935 \pm 0.487$ \\
   \cmidrule{2-9}
   & DPA2DL & $0.574 \pm 0.328$ & $0.630 \pm 0.374$ & $1.367 \pm 0.338$ & $0.793 \pm 0.540$ & $0.395 \pm 0.343$ & $0.458 \pm 0.569$ & $0.703 \pm 0.394$\\
   & DPA2DL$\times2$ & $ 0.884 \pm 0.381$ & $1.069 \pm 0.497$ & $1.333 \pm 0.348$ & $1.451 \pm 0.692$ & $0.809 \pm 0.415$ & $1.004 \pm 0.496$  & $1.092 \pm 0.467$\\
   & DPA2DL+GP & $\mathbf{1.077 \pm 0.476}$ & $\mathbf{1.450 \pm 0.707}$ & $1.493 \pm 0.401$ & $\mathbf{1.699 \pm 0.815}$ & $\mathbf{1.147 \pm 0.585}$ & $\mathbf{1.623 \pm 0.989}$ & $\mathbf{1.415 \pm 0.650}$\\
   \cmidrule{3-9}
   & GP Gain\textsuperscript{2} & $+ 21.83\% \uparrow$ & $+35.64\%\uparrow$ & $+12.00\%\uparrow$ & $+17.09\%\uparrow$ & $+41.77\%\uparrow$ & $+61.65\%\uparrow$ & $+29.57\%\uparrow$\\
   \midrule
   \multirow{10}{*}{Yelp} & None & $0.205 \pm 0.194$ & $0.217 \pm 0.288$ & $0.220 \pm 0.257$ & $0.168 \pm 0.204$ & $0.225 \pm 0.252$ & $0.342 \pm 0.538$ & $0.229 \pm 0.287$\\
   \cmidrule{2-9}
   & Random & $0.099 \pm 0.071$ & $0.053 \pm 0.064$ & $1.529 \pm 0.180$ & $0.156 \pm 0.169$ & $0.347 \pm 0.256$ & $0.260 \pm 0.435$ & $0.407 \pm 0.184$\\
   & Bandwagon & $0.211 \pm 0.140$ & $0.237 \pm 0.120$ & $1.197 \pm 0.210$ & $0.326 \pm 0.138$ & $0.291 \pm 0.231$ & $0.135 \pm 0.216$ & $0.399 \pm 0.166$\\
   \cmidrule{2-9}
   & PGA & $0.245 \pm 0.148$ & $0.461 \pm 0.312$
   & $0.483 \pm 0.313$ & $0.238 \pm 0.209$ & $0.343 \pm 0.227$ & $0.287 \pm 0.444$ & $0.343 \pm 0.271$\\
   \cmidrule{2-9}
   & RAPU-R & $0.193 \pm 0.192$ & $0.448 \pm 0.223$ & $1.193 \pm 0.424$ & $0.529 \pm 0.249$ & $0.269 \pm 0.295$ & $0.198 \pm 0.313$ & $0.472 \pm 0.252$ \\
   & RAPU-R$\times2$ & $0.236 \pm 0.227$ & $0.499 \pm 0.267$ & $1.140 \pm 0.392$ & $0.514 \pm 0.206$ & $0.275 \pm 0.331$ & $0.229 \pm 0.373$ & $0.482 \pm 0.283$ \\
   & RAPU-R+GP & $0.533 \pm 0.226$ & $0.863 \pm 0.309$ & $0.323 \pm 0.161$ & $0.398 \pm 0.246$ & $0.481 \pm 0.300$ & $0.693 \pm 0.559$ & $0.549 \pm 0.296$ \\
   \cmidrule{2-9}
   & DPA2DL & $0.862 \pm 0.202$ & $1.676 \pm 0.188$ & $1.560 \pm 0.431$ & $1.450 \pm 0.338$ & $0.761 \pm 0.389$ & $1.409 \pm 0.579$ & $1.286 \pm 0.329$\\
   & DPA2DL$\times2$ & $0.882 \pm 0.295$ & $1.538 \pm 0.332$ & $1.575 \pm 0.436$ & $1.616 \pm 0.408$ & $0.705 \pm 0.329$ & $1.508 \pm 0.789$  & $1.304 \pm 0.403$\\
   & DPA2DL+GP & $\mathbf{0.992 \pm 0.391}$ & $\mathbf{1.742 \pm 0.697}$ & $\mathbf{1.684 \pm 0.565}$ & $\mathbf{1.786 \pm 0.616}$ & $\mathbf{1.048 \pm 0.607}$ & $\mathbf{1.985 \pm 1.381}$ & $\mathbf{1.539 \pm 0.654}$\\
   \cmidrule{3-9}
   & GP Gain & $+12.47\%\uparrow$ & $+13.26\%\uparrow$ & $+6.92\%\uparrow$ & $+10.51\%\uparrow$ & $+48.65\%\uparrow$ & $+31.63\%\uparrow$ & $+18.02\%\uparrow$\\
   \midrule
   \multirow{8}{*}{Tenrec} & None & $0.014 \pm 0.020$ & $0.014 \pm 0.011$ & $0.027 \pm 0.027$ & $0.010 \pm 0.013$ & {OOM\textsuperscript{3}} & $0.012 \pm 0.013$ & $0.015 \pm 0.017$\\
   \cmidrule{2-9}
   & Random & $0.001 \pm 0.000$ & $0.043 \pm 0.013$ & $\mathbf{0.152 \pm 0.020}$ & $0.198 \pm 0.030$ & {OOM} & $0.284 \pm 0.011$ & $0.135 \pm 0.006$\\
   & Bandwagon & $0.006 \pm 0.003$ & $0.032 \pm 0.012$ & $0.037 \pm 0.006$ & $0.162 \pm 0.022$ & {OOM} & $0.074 \pm 0.014$ & $0.062 \pm 0.005$\\
   \cmidrule{2-9}
    & RAPU-R & $0.004 \pm 0.004$ & $0.009 \pm 0.009$ & $0.011 \pm 0.003$ & $0.061 \pm 0.018$ & {OOM} & $0.007 \pm 0.003$ & $0.018 \pm 0.007$ \\
    & RAPU-R$\times2$ & $0.004 \pm 0.002$ & $0.005 \pm 0.003$ & $0.011 \pm 0.005$ & $0.012 \pm 0.006$ & {OOM} & $0.022 \pm 0.007$ & $0.011 \pm 0.004$ \\
     & RAPU-R+GP & $0.006 \pm 0.004$ & $0.008 \pm 0.010$ & $0.013 \pm 0.007$ & $0.011 \pm 0.003$ & {OOM} & $0.032 \pm 0.015$ & $0.014 \pm 0.006$ \\
   \cmidrule{2-9}
   & DPA2DL & $0.070 \pm 0.010$ & $0.083 \pm 0.019$ & $0.124 \pm 0.022$ & $0.137 \pm 0.042$ & {OOM} & $0.068 \pm 0.007$ & $0.096 \pm 0.015$\\
   & DPA2DL$\times2$ & $0.077 \pm 0.021$ & $0.064 \pm 0.008$ & $0.077 \pm 0.016$ & $0.079 \pm 0.059$ & {OOM} & $0.099 \pm 0.008$ & $0.079 \pm 0.014$\\
   & DPA2DL+GP & $\mathbf{0.099 \pm 0.017}$ & $\mathbf{0.164 \pm 0.036}$ & $0.103 \pm 0.015$ & $\mathbf{0.395 \pm 0.118}$ & {OOM} & $\mathbf{0.335 \pm 0.063}$ & $\mathbf{0.219 \pm 0.015}$\\
   \cmidrule{3-9}
   & GP Gain & $+28.57\%\uparrow$ & $+156.25\%\uparrow$ & $+33.76\%\uparrow$ & $+400.00\%\uparrow$ & {OOM} & $+238.38\%\uparrow$ & $+177.21\%$\\
   \bottomrule
\end{tabular}}
\begin{minipage}{\textwidth}  
        \footnotesize      
        \textsuperscript{1}: Average $\text{Recall}@50 (\%)$ across 6 victim recommender systems. \\
        \textsuperscript{2}: $(\text{Recall}@50_{\text{DPA2DL+GP}} - \text{Recall}@50_{\text{DPA2DL}\times2}) / \text{Recall}@50_{\text{DPA2DL}\times 2}.$ \\
        \textsuperscript{3}: PGA, RevAdv attacker and NeuMF recommender cannot be applied to large-scale datasets, due to GPU memory limit.
\end{minipage}
\end{table*}

\begin{table}[t]
  \caption{Time~(s) Comparison: three DPA2DL variants.}
  \label{tb:atk_time}
  \centering
  \renewcommand{\arraystretch}{0.6}
  \scalebox{0.9}{
  \begin{tabular}{lcccc}
    \toprule
    Dataset & DPA2DL & DPA2DL$\times2$& DPA2DL+GP& Reduction\\
    \midrule
    Gowalla & $885$ & $1761$ & $999$ & $43.27\% \downarrow$ \\
    Yelp & $2730 $ & $5500$ & $3270$ & $40.54\% \downarrow$\\
    DPA2DL & $39358 $ & $72140$ & $52899$ & $26.67\% \downarrow$\\
    \bottomrule
\end{tabular}}
\end{table}

\subsubsection{Evaluation Protocol}
\label{sec:protocol}
We evaluate the effectiveness of poisoning attacks in a black-box context, utilizing a fixed surrogate model to attack multiple victim recommenders. 
The selected surrogate models are MF-MSE for PGA and RevAdv, and MF-BCE for DPA2DL and RAPU-R.
Each attack generates fake user interactions under certain budgets and injects them into the poisoned training ($80\%$) and validation ($20\%$) sets, which are used to retrain victim recommenders from scratch.
For each dataset, we randomly select $5$ items from all items as our target item set and repeat this process $5$ times, following~\cite{ li2022revisiting}. 
The results reported represent the averages and standard deviations across $5$ target item sets.

Because the attack targets a set of items $\mathcal{I}_t$, we quantify the attack performance using Recall, which is defined as:
\begin{equation}
\text{Recall}@k = \frac{1}{\vert \mathcal{U}^r \setminus \mathcal{U}_{\mathcal{I}_t}^{\text{all}} \vert} \sum_{\mathsf{u} \in \mathcal{U}^r \setminus \mathcal{U}_{\mathcal{I}_t}^{\text{all}}} \frac{\vert \mathcal{T}_{\mathsf{u}} \cap \mathcal{I}_t \vert}{\vert \mathcal{I}_t \setminus \mathcal{I}_{\mathsf{u}} \vert}
\end{equation}
where $\mathcal{U}_{\mathcal{I}_t}^{\text{all}}$ denotes the set of users who have interacted with all items in $\mathcal{I}_t$, and $\mathcal{T}_{\mathsf{u}}$ denote the top-$k$ recommendation set for user $\mathsf{u}$. 
The Recall metric simplifies to HR defined in Equation (\ref{eq:hr}), when the target item set $\mathcal{I}_t$ contains only one item.
Consistent with prior research~\cite{tang2020revisiting}, we set $k=50$, i.e., $\vert \mathcal{T}_{\mathsf{u}} \vert=50$.

\subsubsection{Baseline Attack Methods}
The experiments on data poisoning attacks against recommender systems utilize both heuristic (Random, Bandwagon) and optimization-based (PGA, RevAdv, RAPU-R, DPA2DL) attacks as baselines. 

\begin{itemize}[leftmargin=*]
  \item \textbf{None}: This refers to scenarios where no attack is executed.
  \item \textbf{Random Attack}~\cite{lam2004shilling}: In this attack, fake users interact with the target items along with some random items.
  \item \textbf{Bandwagon Attack}~\cite{o2005recommender}: Building upon Random attack, Bandwagon attack additionally involves some popular items.
  \item \textbf{PGA Attack}~\cite{li2016data}: It specifically targets factorization-based recommenders by using an analytic solution of adversarial gradients.
  \item \textbf{RevAdv}~\cite{tang2020revisiting}: This attack computes higher-order adversarial gradients of retraining by automatic differentiation libraries.
  \item \textbf{RAPU-R}~\cite{zhang2021data}: It reverses the optimization process of recommendation models to construct fake user interactions.
  \item \textbf{DPA2DL}~\cite{huang2021data}: This attack simulates a deep learning based poisoned recommender, and generates fake users by its predictions.
\end{itemize}

Optimization-based methods all rely on retraining a surrogate recommender.
Among them, PGA and RevAdv require computing adversarial gradients, which can not scale to large datasets. 
Therefore, we incorporated GP into RAPU-R and DPA2DL to investigate its potential in enhancing poisoning attacks.

\subsubsection{Victim Recommender Systems}
We evaluated the attack effectiveness using five representative CF methods as the victims, including a robust defense method MF-APR.  

\begin{itemize}[leftmargin=*]
  \item \textbf{MF-BPR}:  The most basic latent factor model Matrix Factorization (MF)~\cite{koren2009matrix} optimized by Bayesian Personalized Ranking loss~\cite{Rendle2009BPRBP}.
  \item \textbf{MF-APR}:  This variant of MF utilizes the Adversarial Personalized Ranking framework~\cite{Rendle2009BPRBP}, enhancing the robustness of BPR through adversarial training on embedding parameters.
  \item \textbf{Mult-VAE}~\cite{liang2018variational}: It employs the variational autoencoder architecture~\cite{kingma2013auto} to encode and decode users’ interaction behaviors.
  \item \textbf{NeuMF}~\cite{he2017neural}: It employs MLP to model the nonlinear interactions between the representations of users and items. 
  \item \textbf{LightGCN} \cite{he2020lightgcn}: A state-of-the-art recommendation method, using a simplified version of Graph Convolutional Network (GCN).
  \item \textbf{MF-BCE}: MF trained by BCE loss function, which is the surrogate recommender used in our implemented RAPU-R and DPA2DL.
\end{itemize}

\begin{table*}
  \caption{Data poisoning attack on single unpopular target item, evaluated by $\text{HR}@50 (\%)$.}
  \label{tb:attack_unpopular}
  \scalebox{0.98}{
  \renewcommand{\arraystretch}{0.8}
  \begin{tabular}{ccccccccc}
    \toprule
   Dataset & Attacker & {MF-BPR} & {MF-APR} & {LightGCN} & {MultiVAE} & {NeuMF} & {MF-BCE} & {Average} \\
   \midrule
   \multirow{2}{*}{Gowalla} 
   & DPA2DL$\times2$ & $0.508 \pm 0.092$ & $0.532 \pm 0.140$ & $\mathbf{1.010 \pm 0.172}$ & $1.996 \pm 0.303$ & $0.468 \pm 0.113$ & $0.660 \pm 0.193$ & $0.863 \pm 0.110$ \\
   & DPA2DL+GP & $\mathbf{1.304 \pm 0.464}$ & $\mathbf{1.217 \pm 0.329}$ & $0.969 \pm 0.220$ & $\mathbf{2.262 \pm 0.372}$ & $\mathbf{0.856 \pm 0.160}$ & $\mathbf{0.913 \pm 0.270}$ & $\mathbf{1.253 \pm 0.274}$ \\
   \cmidrule{2-9}
   \multirow{2}{*}{Yelp} 
    & DPA2DL$\times2$ & $1.450 \pm 0.102$ & $2.359 \pm 0.115$ & $1.565 \pm 0.053$ & $2.846 \pm 0.271$ & $1.217 \pm 0.241$ & $1.736 \pm 0.164$ & $1.862 \pm 0.103$ \\
    & DPA2DL+GP & $\mathbf{1.858 \pm 0.168}$ & $\mathbf{2.513 \pm 0.338}$ & $\mathbf{1.712 \pm 0.132}$ & $\mathbf{3.083 \pm 0.195}$ & $\mathbf{1.434 \pm 0.175}$ & $\mathbf{1.900 \pm 0.166}$ & $\mathbf{2.084 \pm 0.110}$ \\
   \bottomrule
\end{tabular}}
\end{table*}

\subsubsection{Parameter Settings}
\label{sec:param_setting}
We implement GP using PyTorch~\cite{paszke2019pytorch}. 
The entire source code, including data preparation, hyper-parameter tuning, baseline attack methods, and victim recommenders, is accessible on GitHub\footnote{\url{https://github.com/WuYunfan/GradientPassingAttack}}. 
For all attack methods and recommenders, we adjust their hyper-parameters for each dataset on the validation set.
Specific to GP, we set the GP layer $l$ to $2$ by default. 
The threshold $\xi$ is tuned across $\{-\infty, 0, \infty\}$, and the weight $\alpha$ across $\{0.1, 1, 10, 100, 1000\}$. 
We introduce a small number of fake users, amounting to $1\%$ of real users, denoted as $n^f=0.01n^r$. The interaction budget $\tau$ is equal to the average number of interactions across real users in each dataset. 

\begin{table*}
  \caption{Attack performances of four DPA2DL variants evaluated by $\text{Recall}@50 (\%)$, when the surrogate is pre-trained.}
  \label{tb:atk_pretrain}
  \scalebox{0.95}{
  \renewcommand{\arraystretch}{0.7}
  \begin{tabular}{ccccccccc}
    \toprule
   Dataset & Attacker & {MF-BPR} & {MF-APR} & {LightGCN} & {MultiVAE} & {NeuMF} & {MF-BCE} & {Average} \\
   \midrule
   \multirow{4}{*}{Gowalla} & Pre-train & $1.039 \pm 0.568$ & $1.250 \pm 0.719$ & $1.376 \pm 0.457$ & $\mathbf{1.438 \pm 0.789}$ & $1.084 \pm 0.830$ & $1.783 \pm 1.516$ & $1.328 \pm 0.804$\\
   & +Sample & $1.779 \pm 0.494$ & $2.522 \pm 0.450$ & $5.188 \pm 2.312$ & $0.850 \pm 0.397$ & $\mathbf{2.863 \pm 0.855}$ & $\mathbf{20.705 \pm 1.786}$ & $5.651 \pm 0.842$\\
   & +Sample$\times2$ & $1.722 \pm 0.443$ & $1.964 \pm 0.504$ & $5.230 \pm 2.120$ & $0.906 \pm 0.464$ & $1.527 \pm 0.670$ & $8.646 \pm 1.745$ & $3.332 \pm 0.799$\\
   & +Sample+GP & $\mathbf{4.540 \pm 1.791}$ & $\mathbf{4.981 \pm 1.691}$ & $\mathbf{14.103 \pm 5.367}$ & $1.091 \pm 0.548$ & $1.870 \pm 0.472$ & $17.193 \pm 2.685$ & $\mathbf{7.296 \pm 1.896}$\\
   \midrule
   \multirow{4}{*}{Yelp} & Pre-train & $0.869 \pm 0.348$ & $1.619 \pm 0.621$ & $1.444 \pm 0.342$ & $\mathbf{1.777 \pm 0.392}$ & $1.139 \pm 0.549$ & $2.650 \pm 1.686$ & $1.583 \pm 0.610$\\
   & +Sample & $1.801 \pm 0.348$ & $2.504 \pm 0.380$ & $11.038 \pm 1.008$ & $0.897 \pm 0.214$ & $1.160 \pm 0.441$ & $14.086 \pm 2.499$ & $5.248 \pm 0.594$\\
   & +Sample$\times2$ & $1.223 \pm 0.357$ & $1.652 \pm 0.483$ & $11.631 \pm 1.045$ & $0.801 \pm 0.327$ & $1.031 \pm 0.385$ & $9.759 \pm 2.608$ & $4.349 \pm 0.696$\\
   & +Sample+GP & $\mathbf{8.024 \pm 0.914}$ & $\mathbf{11.703 \pm 0.854}$ & $\mathbf{13.132 \pm 0.960}$ & $1.557 \pm 0.753$ & $\mathbf{2.420 \pm 0.639}$ & $\mathbf{21.190 \pm 3.480}$ & $\mathbf{9.671 \pm 0.767}$\\
   \bottomrule
\end{tabular}}
\end{table*}

\subsection{Enhancing Poisoning Attacks (Q1)}
This section investigates black-box poisoning attacks, with detailed attack settings described in Sections \ref{sec:protocol} and \ref{sec:param_setting}. 
Notably, hyper-parameters of attackers, including surrogate learning rate and $\ell_2$ regularization, are optimized before applying GP. 
To control the time costs, we limit the number of retraining epochs to $1$ for RAPU-R, DPA2DL, and their GP-enhanced variants. 
The number of training iterations depends on dataset size and batch size. 
For Yelp and Tenrec, GP is applied with probabilities of $0.5$ and $0.25$, respectively, rather than at every iteration. 
Additionally, we evaluate RAPU-R$\times2$ and DPA2DL$\times2$ with $2$ retraining epochs to assess GP's efficiency.

Table~\ref{tb:attack} presents the results using $\text{Recall}@50 (\%) = \text{Recall}@50 \times 100$ as the primary metric, while Table~\ref{tb:atk_time} compares the time costs of three DPA2DL variants.
Among the baselines, DPA2DL emerges as the most scalable and advanced method. 
DPA2DL$\times2$ generally outperforms DPA2DL, particularly when the victim model is MF-BCE, which aligns with its surrogate model. 
This suggests that a more accurate surrogate model, closely approximating the victim recommender, yields a more potent attack.

DPA2DL+GP is the most effective attack method, primarily due to GP's ability to accelerate iterative surrogate retraining and provide more accurate feedback for optimizing fake users. 
Specifically, GP improves the average attack effectiveness by $29.57\%$, $18.02\%$, and $177.21\%$ across three datasets while reducing attack time by $43.27\%$, $40.54\%$, and $26.67\%$ when comparing DPA2DL+GP with DPA2DL$\times2$. 
By assigning distinct weights and thresholds for odd and even terms, GP strategically focuses on influential passing paths, surpassing the effects of simply doubling retraining epochs.
RAPU-R exhibits similar results to DPA2DL when enhanced by GP, except for its failure on the large Tenrec dataset. 

Attack results against LightGCN differ from other victims, possibly due to its graph architecture inadvertently facilitating the propagation of attack influence. 
However, as none of the baselines consider it as the surrogate recommender, transferability to LightGCN is not guaranteed. 
The robust framework APR does not consistently enhance resistance against attacks, because it targets parameter perturbation attacks rather than data poisoning. 
Large standard deviations in Table~\ref{tb:attack} are attributed to the diverse characteristics of random target item sets. 
Table~\ref{tb:attack_unpopular} provides additional experiments targeting a single unpopular item, further demonstrating GP's effectiveness in promoting less-favored items.

\subsection{Generalizability in Pre-training Setting (Q2)}
The surrogate recommender is repeatedly retrained on the poisoned dataset with the latest fake users. 
However, the majority of the poisoned data, i.e., real user interactions, remains unchanged.
Therefore, it is feasible to pre-train a surrogate recommender using only real interactions.
Then the parameter weights from the pre-trained surrogate can be used to initialize the iterative surrogate retraining during fake users' optimization.
It is expected to yield a better surrogate recommender under limited retraining epochs.

We investigate integrating a pre-trained surrogate recommender with the sampling strategy inspired by incremental learning.
Specifically, a surrogate is first pre-trained on real interactions with sufficient epochs.
During each retraining, a sampled poisoned dataset is constructed, comprising $10\%$ randomly sampled real interactions and all fake ones, emphasizing the attack impact of fake users. 
The effectiveness and generalizability of GP are assessed when combined with these techniques.
Four variants of DPA2DL are compared: \textbf{Pre-train}, \textbf{Pre-train+Sampling}, \textbf{Pre-train+Sampling$\times2$} and \textbf{Pre-train+Sampling+GP}.

The outcomes are summarized in Table~\ref{tb:atk_pretrain}.
A comparison between Pre-train and original DPA2DL (Table~\ref{tb:attack}) shows the efficacy of pre-training a surrogate recommender beforehand.
The combination of pre-training and sampling further enhances the attack significantly.
However, doubling the retraining epochs decreases the performance when comparing Pre-train+Sampling$\times2$ with Pre-train+Sampling.
It may be attributed to potential over-fitting to the biased sampled dataset, as it only contains partial real interactions. 
Consequently, over-training on it may reduce the accuracy of the surrogate.
GP alleviates this by training on the sampled dataset while passing gradients on the whole dataset, further improving the attack performance.
While the strategy of constructing a sampled dataset may introduce the over-fitting issue, it is effective in many cases, underlining the under-explored potential of incremental learning techniques in enhancing poisoning attacks.

\subsection{More Analyses (Q3)}
\subsubsection{Gradient Similarity between Interacted User-item Pairs}
Our proposed GP is primarily driven by the intuition that gradients between interacted user-item pairs show high similarity during a period. 
Thus, explicitly passing gradients within every training iteration can bring additional optimization signals for users and items, accelerating the surrogate retraining. 
To further support this hypothesis, we compute the cosine similarity of average gradients aggregated over one epoch for interacted user-item pairs. 
The mean and standard deviation across all pairs are recorded. For comparison, we also select an equal number of random user-item pairs.

\begin{figure}[t]
  \centering
  \includegraphics[width=\linewidth]{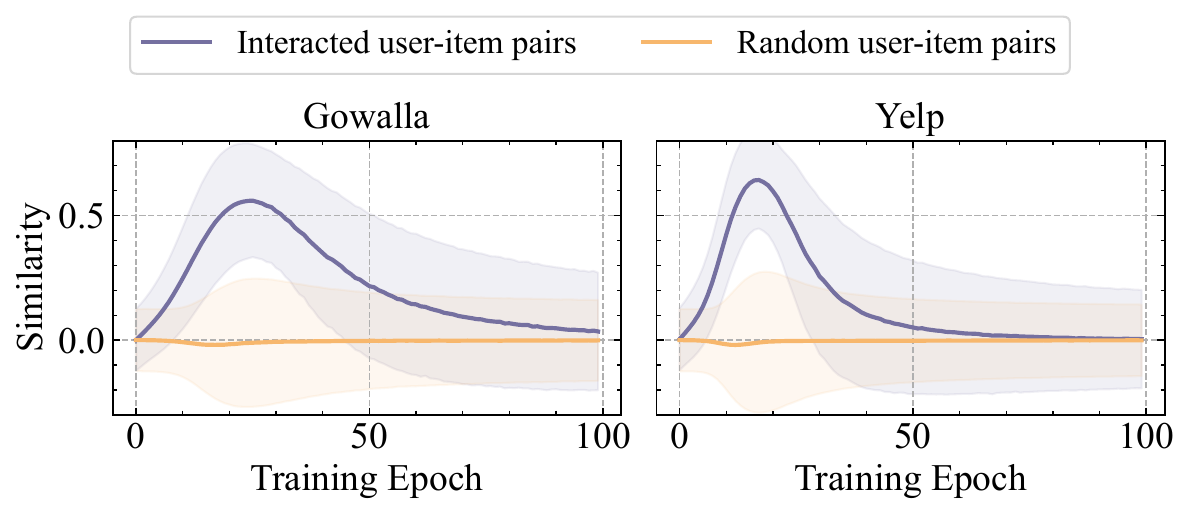}
  \caption{Comparison of gradient similarity between interacted and random user-item pairs.}
  \label{fig:grad_sim}
\end{figure}

Figure \ref{fig:grad_sim} shows the outcomes on Gowalla and Yelp datasets. 
There is a clear difference between the similarity of interacted pairs and random ones. 
During early training, gradient similarity among interacted pairs initially rises exceeding $0.5$, then diminishes to $0$. 
This is because the representations of users and items start with random initialization at epoch $0$ and undergo optimization to find optimal positions in the embedding space. 
So similarity increases during this optimization process. 
Towards the end of training, all gradients have small values with decreased similarity.

\subsubsection{Retraining Enhancement of GP}
Retraining the surrogate recommender is crucial in data poisoning attacks.
To demonstrate the effectiveness of GP in accelerating retraining and obtaining a more accurate surrogate, we conduct experiments on Gowalla to evaluate the surrogate's capability in replicating the victim's behavior.
Two CF methods, MF and LightGCN, along with two loss functions, BPR and BCE, are employed in the experiments.

For each experiment, a victim recommender is first trained over $1000$ epochs. 
We then train a surrogate recommender with the same architecture, but under different epoch constraints ($1,5,10,50,100$).
The similarity between the recommendation lists of surrogate and victim recommenders is evaluated by Jaccard Index~\cite{costa2021further} averaged across all users. 
The surrogates trained with and without GP are compared, using their optimal learning rate and $\ell_2$ regularization specifically for each experiment.

\begin{figure}[t]
  \centering
  \includegraphics[width=\linewidth]{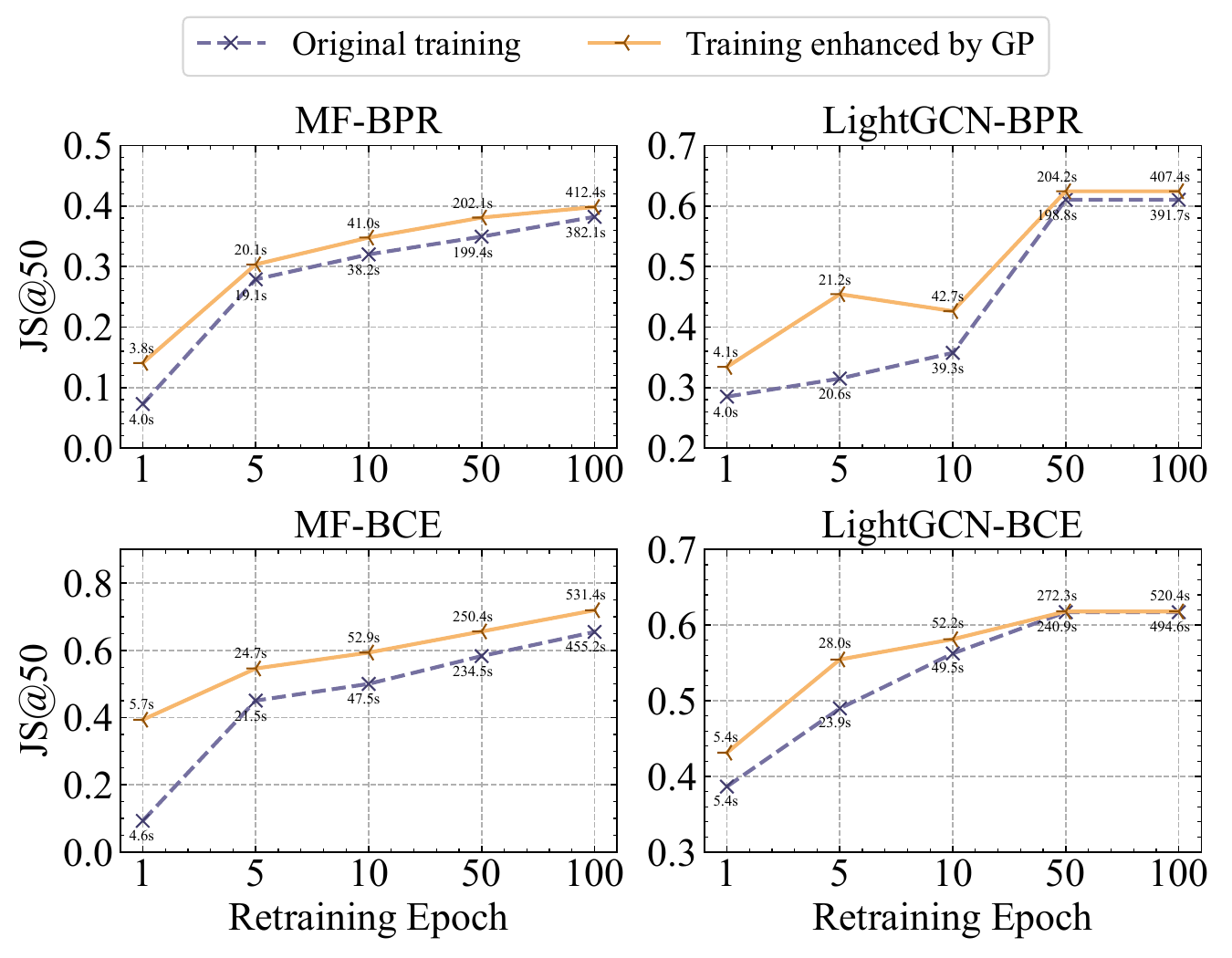}
  \caption{Jaccard Similarity between the surrogate and victim recommenders across various retraining epochs, on Gowalla.}
  \label{fig:retrain_gowalla}
\end{figure}

The results of retraining are illustrated in Figure \ref{fig:retrain_gowalla} with consumed time annotated by texts. 
Surrogate recommenders trained with GP consistently achieve higher similarity than the original ones under same epochs, demonstrating the effectiveness and generalizability of GP across different models and loss functions. 
It underscores the capability of GP to enhance the behavioral similarity between the surrogate and the victim recommender, potentially leading to stronger attacks. 
Moreover, training with GP for just $5$ epochs attains results comparable to or even surpassing the original training for $10$ epochs, highlighting the efficiency of GP. 

\subsubsection{Hyper-parameter Analysis}
\begin{figure}[t]
  \centering
  \includegraphics[width=\linewidth]{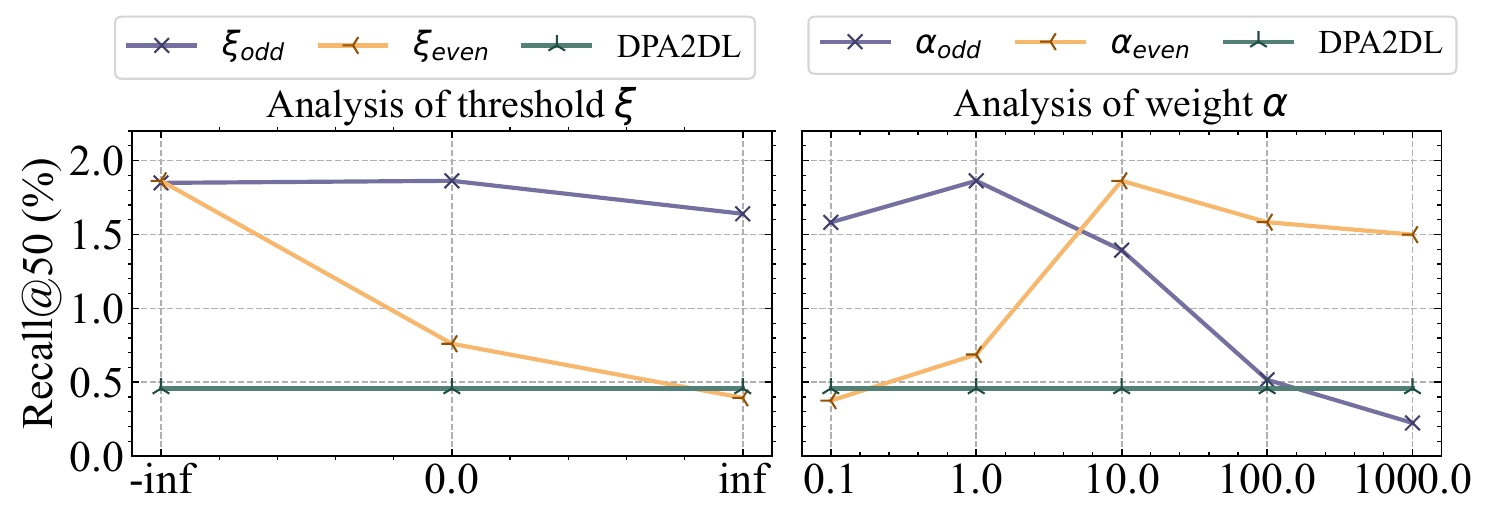}
  \caption{Hyperparameter Analysis on Gowalla.}
  \label{fig:hyper}
\end{figure}

We conduct experiments to investigate the effectiveness of GP with different hyper-parameters. 
The results attacking MF-BCE recommender on Gowalla dataset are presented in Figure \ref{fig:hyper}.
By default, $\xi_{\text{odd}}$ and $\xi_{\text{even}}$ are set to $0$ and $-\infty$, respectively, while $\alpha_{\text{odd}}$ and $\alpha_{\text{even}}$ are set to $1$ and $10$.

The optimal thresholds $\xi$ and weights $\alpha$ for odd and even terms differ, confirming the validity of our GP design. 
The even terms related to \emph{GP within user pairs and item pairs}, have a more significant impact. 
It suggests that the gradients of users may exhibit some incongruity with those of items, leading users to prefer adopting gradients from other users rather than items. 
Furthermore, GP surpasses DPA2DL across most hyper-parameter configurations.

\section{Conclusion}
We propose Gradient Passing (GP), a novel technique that accelerates the surrogate retraining in poisoning attacks by passing gradients between the representations of interacted user-item pairs during backpropagation. 
Through theoretical analysis and extensive experiments on real-world datasets, we demonstrate that GP can significantly accelerate the retraining process. 
When integrated into existing attack methods, GP improves their attack effectiveness by enabling a closer approximation of the surrogate recommender to the victim and providing better attack feedback for optimizing fake users. 
Since most optimization-based poisoning attacks require the time-consuming surrogate retraining, GP provides a simple yet effective solution to enhance attacks against recommender systems. 
As for defense, an effective approach to mitigate the risk is preventing the leakage of interaction data to potential attackers. 
By securing user data, we can significantly reduce the effectiveness of such poisoning attacks. 
Further research can explore the potential of GP in enhancing general training of recommenders and extend GP from CF to other tasks like sequential recommendation.

\begin{acks}
This work is funded by the Strategic Priority Research Program of the Chinese Academy of Sciences under Grant No. XDB0680101, and the National Natural Science Foundation of China under Grant Nos. 62272125, 62102402, U21B2046. Huawei Shen is also supported by Beijing Academy of Artificial Intelligence (BAAI).
\end{acks}
\bibliographystyle{ACM-Reference-Format}
\balance
\bibliography{base}


\begin{thebibliography}{57}


\ifx \showCODEN    \undefined \def \showCODEN     #1{\unskip}     \fi
\ifx \showDOI      \undefined \def \showDOI       #1{#1}\fi
\ifx \showISBNx    \undefined \def \showISBNx     #1{\unskip}     \fi
\ifx \showISBNxiii \undefined \def \showISBNxiii  #1{\unskip}     \fi
\ifx \showISSN     \undefined \def \showISSN      #1{\unskip}     \fi
\ifx \showLCCN     \undefined \def \showLCCN      #1{\unskip}     \fi
\ifx \shownote     \undefined \def \shownote      #1{#1}          \fi
\ifx \showarticletitle \undefined \def \showarticletitle #1{#1}   \fi
\ifx \showURL      \undefined \def \showURL       {\relax}        \fi
\providecommand\bibfield[2]{#2}
\providecommand\bibinfo[2]{#2}
\providecommand\natexlab[1]{#1}
\providecommand\showeprint[2][]{arXiv:#2}

\bibitem[Burke et~al\mbox{.}(2005)]%
        {burke2005limited}
\bibfield{author}{\bibinfo{person}{Robin Burke}, \bibinfo{person}{Bamshad Mobasher}, {and} \bibinfo{person}{Runa Bhaumik}.} \bibinfo{year}{2005}\natexlab{}.
\newblock \showarticletitle{Limited knowledge shilling attacks in collaborative filtering systems}. In \bibinfo{booktitle}{\emph{Proceedings of 3rd international workshop on intelligent techniques for web personalization (ITWP 2005), 19th international joint conference on artificial intelligence (IJCAI 2005)}}. \bibinfo{pages}{17--24}.
\newblock


\bibitem[Chen et~al\mbox{.}(2022)]%
        {chen2022knowledge}
\bibfield{author}{\bibinfo{person}{Jingfan Chen}, \bibinfo{person}{Wenqi Fan}, \bibinfo{person}{Guanghui Zhu}, \bibinfo{person}{Xiangyu Zhao}, \bibinfo{person}{Chunfeng Yuan}, \bibinfo{person}{Qing Li}, {and} \bibinfo{person}{Yihua Huang}.} \bibinfo{year}{2022}\natexlab{}.
\newblock \showarticletitle{Knowledge-enhanced Black-box Attacks for Recommendations}. In \bibinfo{booktitle}{\emph{Proceedings of the 28th ACM SIGKDD Conference on Knowledge Discovery and Data Mining}}. \bibinfo{pages}{108--117}.
\newblock


\bibitem[Chen et~al\mbox{.}(2023)]%
        {chen2023dark}
\bibfield{author}{\bibinfo{person}{Ziheng Chen}, \bibinfo{person}{Fabrizio Silvestri}, \bibinfo{person}{Jia Wang}, \bibinfo{person}{Yongfeng Zhang}, {and} \bibinfo{person}{Gabriele Tolomei}.} \bibinfo{year}{2023}\natexlab{}.
\newblock \showarticletitle{The dark side of explanations: Poisoning recommender systems with counterfactual examples}.
\newblock \bibinfo{journal}{\emph{arXiv preprint arXiv:2305.00574}} (\bibinfo{year}{2023}).
\newblock


\bibitem[Cho et~al\mbox{.}(2011)]%
        {cho2011friendship}
\bibfield{author}{\bibinfo{person}{Eunjoon Cho}, \bibinfo{person}{Seth~A Myers}, {and} \bibinfo{person}{Jure Leskovec}.} \bibinfo{year}{2011}\natexlab{}.
\newblock \showarticletitle{Friendship and mobility: user movement in location-based social networks}. In \bibinfo{booktitle}{\emph{Proceedings of the 17th ACM SIGKDD International Conference on Knowledge Discovery \& Data Mining}} \emph{(\bibinfo{series}{KDD '11})}. \bibinfo{pages}{1082--1090}.
\newblock


\bibitem[Christakopoulou and Banerjee(2019)]%
        {christakopoulou2019adversarial}
\bibfield{author}{\bibinfo{person}{Konstantina Christakopoulou} {and} \bibinfo{person}{Arindam Banerjee}.} \bibinfo{year}{2019}\natexlab{}.
\newblock \showarticletitle{Adversarial attacks on an oblivious recommender}. In \bibinfo{booktitle}{\emph{Proceedings of the 13th ACM Conference on Recommender Systems}}. \bibinfo{pages}{322--330}.
\newblock


\bibitem[Costa(2021)]%
        {costa2021further}
\bibfield{author}{\bibinfo{person}{Luciano da~F Costa}.} \bibinfo{year}{2021}\natexlab{}.
\newblock \showarticletitle{Further generalizations of the Jaccard index}.
\newblock \bibinfo{journal}{\emph{arXiv preprint arXiv:2110.09619}} (\bibinfo{year}{2021}).
\newblock


\bibitem[Covington et~al\mbox{.}(2016)]%
        {covington2016deep}
\bibfield{author}{\bibinfo{person}{Paul Covington}, \bibinfo{person}{Jay Adams}, {and} \bibinfo{person}{Emre Sargin}.} \bibinfo{year}{2016}\natexlab{}.
\newblock \showarticletitle{Deep Neural Networks for YouTube Recommendations}. In \bibinfo{booktitle}{\emph{Proceedings of the 10th ACM Conference on Recommender Systems}} \emph{(\bibinfo{series}{RecSys '16})}. \bibinfo{pages}{191--198}.
\newblock


\bibitem[Diaz-Aviles et~al\mbox{.}(2012)]%
        {diaz2012real}
\bibfield{author}{\bibinfo{person}{Ernesto Diaz-Aviles}, \bibinfo{person}{Lucas Drumond}, \bibinfo{person}{Lars Schmidt-Thieme}, {and} \bibinfo{person}{Wolfgang Nejdl}.} \bibinfo{year}{2012}\natexlab{}.
\newblock \showarticletitle{Real-time top-n recommendation in social streams}. In \bibinfo{booktitle}{\emph{Proceedings of the sixth ACM conference on Recommender systems}}. \bibinfo{pages}{59--66}.
\newblock


\bibitem[Fan et~al\mbox{.}(2021)]%
        {fan2021attacking}
\bibfield{author}{\bibinfo{person}{Wenqi Fan}, \bibinfo{person}{Tyler Derr}, \bibinfo{person}{Xiangyu Zhao}, \bibinfo{person}{Yao Ma}, \bibinfo{person}{Hui Liu}, \bibinfo{person}{Jianping Wang}, \bibinfo{person}{Jiliang Tang}, {and} \bibinfo{person}{Qing Li}.} \bibinfo{year}{2021}\natexlab{}.
\newblock \showarticletitle{Attacking black-box recommendations via copying cross-domain user profiles}. In \bibinfo{booktitle}{\emph{2021 IEEE 37th International Conference on Data Engineering (ICDE)}}. IEEE, \bibinfo{pages}{1583--1594}.
\newblock


\bibitem[Fang et~al\mbox{.}(2020)]%
        {fang2020influence}
\bibfield{author}{\bibinfo{person}{Minghong Fang}, \bibinfo{person}{Neil~Zhenqiang Gong}, {and} \bibinfo{person}{Jia Liu}.} \bibinfo{year}{2020}\natexlab{}.
\newblock \showarticletitle{Influence function based data poisoning attacks to top-n recommender systems}. In \bibinfo{booktitle}{\emph{Proceedings of The Web Conference 2020}}. \bibinfo{pages}{3019--3025}.
\newblock


\bibitem[Fang et~al\mbox{.}(2018)]%
        {fang2018poisoning}
\bibfield{author}{\bibinfo{person}{Minghong Fang}, \bibinfo{person}{Guolei Yang}, \bibinfo{person}{Neil~Zhenqiang Gong}, {and} \bibinfo{person}{Jia Liu}.} \bibinfo{year}{2018}\natexlab{}.
\newblock \showarticletitle{Poisoning attacks to graph-based recommender systems}. In \bibinfo{booktitle}{\emph{Proceedings of the 34th annual computer security applications conference}}. \bibinfo{pages}{381--392}.
\newblock


\bibitem[Goldberg et~al\mbox{.}(1992)]%
        {goldberg1992using}
\bibfield{author}{\bibinfo{person}{David Goldberg}, \bibinfo{person}{David Nichols}, \bibinfo{person}{Brian~M Oki}, {and} \bibinfo{person}{Douglas Terry}.} \bibinfo{year}{1992}\natexlab{}.
\newblock \showarticletitle{Using Collaborative Filtering to Weave an Information Tapestry}.
\newblock \bibinfo{journal}{\emph{Commun. ACM}}  \bibinfo{volume}{35} (\bibinfo{year}{1992}), \bibinfo{pages}{61--70}.
\newblock


\bibitem[He et~al\mbox{.}(2020)]%
        {he2020lightgcn}
\bibfield{author}{\bibinfo{person}{Xiangnan He}, \bibinfo{person}{Kuan Deng}, \bibinfo{person}{Xiang Wang}, \bibinfo{person}{Yan Li}, \bibinfo{person}{Yongdong Zhang}, {and} \bibinfo{person}{Meng Wang}.} \bibinfo{year}{2020}\natexlab{}.
\newblock \showarticletitle{Lightgcn: Simplifying and powering graph convolution network for recommendation}. In \bibinfo{booktitle}{\emph{Proceedings of the 43rd International ACM SIGIR conference on research and development in Information Retrieval}}. \bibinfo{pages}{639--648}.
\newblock


\bibitem[He et~al\mbox{.}(2018)]%
        {he2018outer}
\bibfield{author}{\bibinfo{person}{Xiangnan He}, \bibinfo{person}{Xiaoyu Du}, \bibinfo{person}{Xiang Wang}, \bibinfo{person}{Feng Tian}, \bibinfo{person}{Jinhui Tang}, {and} \bibinfo{person}{Tat-Seng Chua}.} \bibinfo{year}{2018}\natexlab{}.
\newblock \showarticletitle{Outer product-based neural collaborative filtering}. In \bibinfo{booktitle}{\emph{Proceedings of the 27th International Joint Conference on Artificial Intelligence}}. \bibinfo{pages}{2227--2233}.
\newblock


\bibitem[He et~al\mbox{.}(2017)]%
        {he2017neural}
\bibfield{author}{\bibinfo{person}{Xiangnan He}, \bibinfo{person}{Lizi Liao}, \bibinfo{person}{Hanwang Zhang}, \bibinfo{person}{Liqiang Nie}, \bibinfo{person}{Xia Hu}, {and} \bibinfo{person}{Tat-Seng Chua}.} \bibinfo{year}{2017}\natexlab{}.
\newblock \showarticletitle{Neural collaborative filtering}. In \bibinfo{booktitle}{\emph{Proceedings of the 26th international conference on world wide web}}. \bibinfo{pages}{173--182}.
\newblock


\bibitem[Hofmann(2004)]%
        {hofmann2004latent}
\bibfield{author}{\bibinfo{person}{Thomas Hofmann}.} \bibinfo{year}{2004}\natexlab{}.
\newblock \showarticletitle{Latent semantic models for collaborative filtering}.
\newblock \bibinfo{journal}{\emph{ACM Transactions on Information Systems (TOIS)}} \bibinfo{volume}{22}, \bibinfo{number}{1} (\bibinfo{year}{2004}), \bibinfo{pages}{89--115}.
\newblock


\bibitem[Huang and Li(2023)]%
        {huang2023single}
\bibfield{author}{\bibinfo{person}{Chengzhi Huang} {and} \bibinfo{person}{Hui Li}.} \bibinfo{year}{2023}\natexlab{}.
\newblock \showarticletitle{Single-User Injection for Invisible Shilling Attack against Recommender Systems}. In \bibinfo{booktitle}{\emph{Proceedings of the 32nd ACM International Conference on Information and Knowledge Management}}. \bibinfo{pages}{864--873}.
\newblock


\bibitem[Huang et~al\mbox{.}(2021)]%
        {huang2021data}
\bibfield{author}{\bibinfo{person}{Hai Huang}, \bibinfo{person}{Jiaming Mu}, \bibinfo{person}{Neil~Zhenqiang Gong}, \bibinfo{person}{Qi Li}, \bibinfo{person}{Bin Liu}, {and} \bibinfo{person}{Mingwei Xu}.} \bibinfo{year}{2021}\natexlab{}.
\newblock \showarticletitle{Data poisoning attacks to deep learning based recommender systems}. In \bibinfo{booktitle}{\emph{NDSS}}.
\newblock


\bibitem[Kingma and Welling(2013)]%
        {kingma2013auto}
\bibfield{author}{\bibinfo{person}{Diederik~P Kingma} {and} \bibinfo{person}{Max Welling}.} \bibinfo{year}{2013}\natexlab{}.
\newblock \showarticletitle{Auto-encoding variational bayes}.
\newblock \bibinfo{journal}{\emph{arXiv preprint arXiv:1312.6114}} (\bibinfo{year}{2013}).
\newblock


\bibitem[Kipf and Welling(2016)]%
        {kipf2016semi}
\bibfield{author}{\bibinfo{person}{Thomas~N Kipf} {and} \bibinfo{person}{Max Welling}.} \bibinfo{year}{2016}\natexlab{}.
\newblock \showarticletitle{Semi-supervised classification with graph convolutional networks}.
\newblock \bibinfo{journal}{\emph{arXiv preprint arXiv:1609.02907}} (\bibinfo{year}{2016}).
\newblock


\bibitem[Koh and Liang(2017)]%
        {koh2017understanding}
\bibfield{author}{\bibinfo{person}{Pang~Wei Koh} {and} \bibinfo{person}{Percy Liang}.} \bibinfo{year}{2017}\natexlab{}.
\newblock \showarticletitle{Understanding black-box predictions via influence functions}. In \bibinfo{booktitle}{\emph{International conference on machine learning}}. PMLR, \bibinfo{pages}{1885--1894}.
\newblock


\bibitem[Koren et~al\mbox{.}(2009)]%
        {koren2009matrix}
\bibfield{author}{\bibinfo{person}{Yehuda Koren}, \bibinfo{person}{Robert Bell}, {and} \bibinfo{person}{Chris Volinsky}.} \bibinfo{year}{2009}\natexlab{}.
\newblock \showarticletitle{Matrix factorization techniques for recommender systems}.
\newblock \bibinfo{journal}{\emph{Computer}} \bibinfo{volume}{42}, \bibinfo{number}{8} (\bibinfo{year}{2009}), \bibinfo{pages}{30--37}.
\newblock


\bibitem[Lam and Riedl(2004)]%
        {lam2004shilling}
\bibfield{author}{\bibinfo{person}{Shyong~K Lam} {and} \bibinfo{person}{John Riedl}.} \bibinfo{year}{2004}\natexlab{}.
\newblock \showarticletitle{Shilling recommender systems for fun and profit}. In \bibinfo{booktitle}{\emph{Proceedings of the 13th international conference on World Wide Web}}. \bibinfo{pages}{393--402}.
\newblock


\bibitem[Li et~al\mbox{.}(2016)]%
        {li2016data}
\bibfield{author}{\bibinfo{person}{Bo Li}, \bibinfo{person}{Yining Wang}, \bibinfo{person}{Aarti Singh}, {and} \bibinfo{person}{Yevgeniy Vorobeychik}.} \bibinfo{year}{2016}\natexlab{}.
\newblock \showarticletitle{Data poisoning attacks on factorization-based collaborative filtering}.
\newblock \bibinfo{journal}{\emph{Advances in neural information processing systems}}  \bibinfo{volume}{29} (\bibinfo{year}{2016}).
\newblock


\bibitem[LI et~al\mbox{.}(2022)]%
        {li2022revisiting}
\bibfield{author}{\bibinfo{person}{Haoyang LI}, \bibinfo{person}{Shimin DI}, {and} \bibinfo{person}{Lei Chen}.} \bibinfo{year}{2022}\natexlab{}.
\newblock \showarticletitle{Revisiting Injective Attacks on Recommender Systems}.
\newblock \bibinfo{journal}{\emph{Advances in Neural Information Processing Systems}}  \bibinfo{volume}{35} (\bibinfo{year}{2022}), \bibinfo{pages}{29989--30002}.
\newblock


\bibitem[Liang et~al\mbox{.}(2018)]%
        {liang2018variational}
\bibfield{author}{\bibinfo{person}{Dawen Liang}, \bibinfo{person}{Rahul~G Krishnan}, \bibinfo{person}{Matthew~D Hoffman}, {and} \bibinfo{person}{Tony Jebara}.} \bibinfo{year}{2018}\natexlab{}.
\newblock \showarticletitle{Variational autoencoders for collaborative filtering}. In \bibinfo{booktitle}{\emph{Proceedings of the 2018 world wide web conference}}. \bibinfo{pages}{689--698}.
\newblock


\bibitem[Lin et~al\mbox{.}(2020)]%
        {lin2020attacking}
\bibfield{author}{\bibinfo{person}{Chen Lin}, \bibinfo{person}{Si Chen}, \bibinfo{person}{Hui Li}, \bibinfo{person}{Yanghua Xiao}, \bibinfo{person}{Lianyun Li}, {and} \bibinfo{person}{Qian Yang}.} \bibinfo{year}{2020}\natexlab{}.
\newblock \showarticletitle{Attacking recommender systems with augmented user profiles}. In \bibinfo{booktitle}{\emph{Proceedings of the 29th ACM international conference on information \& knowledge management}}. \bibinfo{pages}{855--864}.
\newblock


\bibitem[Mnih and Salakhutdinov(2007)]%
        {mnih2007probabilistic}
\bibfield{author}{\bibinfo{person}{Andriy Mnih} {and} \bibinfo{person}{Russ~R Salakhutdinov}.} \bibinfo{year}{2007}\natexlab{}.
\newblock \showarticletitle{Probabilistic matrix factorization}.
\newblock \bibinfo{journal}{\emph{Advances in neural information processing systems}}  \bibinfo{volume}{20} (\bibinfo{year}{2007}).
\newblock


\bibitem[Mobasher et~al\mbox{.}(2005)]%
        {mobasher2005effective}
\bibfield{author}{\bibinfo{person}{Bamshad Mobasher}, \bibinfo{person}{Robin Burke}, \bibinfo{person}{Runa Bhaumik}, {and} \bibinfo{person}{Chad Williams}.} \bibinfo{year}{2005}\natexlab{}.
\newblock \showarticletitle{Effective attack models for shilling item-based collaborative filtering systems}. In \bibinfo{booktitle}{\emph{Proceedings of the 2005 WebKDD Workshop, held in conjuction with ACM SIGKDD}}, Vol.~\bibinfo{volume}{2005}.
\newblock


\bibitem[Mobasher et~al\mbox{.}(2007)]%
        {mobasher2007toward}
\bibfield{author}{\bibinfo{person}{Bamshad Mobasher}, \bibinfo{person}{Robin Burke}, \bibinfo{person}{Runa Bhaumik}, {and} \bibinfo{person}{Chad Williams}.} \bibinfo{year}{2007}\natexlab{}.
\newblock \showarticletitle{Toward trustworthy recommender systems: An analysis of attack models and algorithm robustness}.
\newblock \bibinfo{journal}{\emph{ACM Transactions on Internet Technology (TOIT)}} \bibinfo{volume}{7}, \bibinfo{number}{4} (\bibinfo{year}{2007}), \bibinfo{pages}{23--es}.
\newblock


\bibitem[O'Mahony et~al\mbox{.}(2005)]%
        {o2005recommender}
\bibfield{author}{\bibinfo{person}{Michael~P O'Mahony}, \bibinfo{person}{Neil~J Hurley}, {and} \bibinfo{person}{Gu{\'e}nol{\'e}~CM Silvestre}.} \bibinfo{year}{2005}\natexlab{}.
\newblock \showarticletitle{Recommender systems: attack types and strategies}. In \bibinfo{booktitle}{\emph{Proceedings of the 20th national conference on Artificial intelligence-Volume 1}}. \bibinfo{pages}{334--339}.
\newblock


\bibitem[Paszke et~al\mbox{.}(2019)]%
        {paszke2019pytorch}
\bibfield{author}{\bibinfo{person}{Adam Paszke}, \bibinfo{person}{Sam Gross}, \bibinfo{person}{Francisco Massa}, \bibinfo{person}{Adam Lerer}, \bibinfo{person}{James Bradbury}, \bibinfo{person}{Gregory Chanan}, \bibinfo{person}{Trevor Killeen}, \bibinfo{person}{Zeming Lin}, \bibinfo{person}{Natalia Gimelshein}, \bibinfo{person}{Luca Antiga}, {et~al\mbox{.}}} \bibinfo{year}{2019}\natexlab{}.
\newblock \showarticletitle{Pytorch: An imperative style, high-performance deep learning library}.
\newblock \bibinfo{journal}{\emph{Advances in neural information processing systems}}  \bibinfo{volume}{32} (\bibinfo{year}{2019}).
\newblock


\bibitem[Qian et~al\mbox{.}(2023)]%
        {qian2023enhancing}
\bibfield{author}{\bibinfo{person}{Fulan Qian}, \bibinfo{person}{Bei Yuan}, \bibinfo{person}{Hai Chen}, \bibinfo{person}{Jie Chen}, \bibinfo{person}{Defu Lian}, {and} \bibinfo{person}{Shu Zhao}.} \bibinfo{year}{2023}\natexlab{}.
\newblock \showarticletitle{Enhancing the Transferability of Adversarial Examples Based on Nesterov Momentum for Recommendation Systems}.
\newblock \bibinfo{journal}{\emph{IEEE Transactions on Big Data}} (\bibinfo{year}{2023}).
\newblock


\bibitem[Rendle et~al\mbox{.}(2009)]%
        {Rendle2009BPRBP}
\bibfield{author}{\bibinfo{person}{S. Rendle}, \bibinfo{person}{C. Freudenthaler}, \bibinfo{person}{Zeno Gantner}, {and} \bibinfo{person}{L. Schmidt-Thieme}.} \bibinfo{year}{2009}\natexlab{}.
\newblock \showarticletitle{BPR: Bayesian Personalized Ranking from Implicit Feedback}. In \bibinfo{booktitle}{\emph{Proceedings of the 25th Conference on Uncertainty in Artificial Intelligence}} \emph{(\bibinfo{series}{UAI '09})}.
\newblock


\bibitem[Rendle et~al\mbox{.}(2020)]%
        {rendle2020neural}
\bibfield{author}{\bibinfo{person}{Steffen Rendle}, \bibinfo{person}{Walid Krichene}, \bibinfo{person}{Li Zhang}, {and} \bibinfo{person}{John Anderson}.} \bibinfo{year}{2020}\natexlab{}.
\newblock \showarticletitle{Neural collaborative filtering vs. matrix factorization revisited}. In \bibinfo{booktitle}{\emph{Proceedings of the 14th ACM Conference on Recommender Systems}}. \bibinfo{pages}{240--248}.
\newblock


\bibitem[Resnick et~al\mbox{.}(1994)]%
        {resnick1994grouplens}
\bibfield{author}{\bibinfo{person}{Paul Resnick}, \bibinfo{person}{Neophytos Iacovou}, \bibinfo{person}{Mitesh Suchak}, \bibinfo{person}{Peter Bergstrom}, {and} \bibinfo{person}{John Riedl}.} \bibinfo{year}{1994}\natexlab{}.
\newblock \showarticletitle{Grouplens: An open architecture for collaborative filtering of netnews}. In \bibinfo{booktitle}{\emph{Proceedings of the 1994 ACM conference on Computer supported cooperative work}}. \bibinfo{pages}{175--186}.
\newblock


\bibitem[Sarwar et~al\mbox{.}(2001)]%
        {sarwar2001item}
\bibfield{author}{\bibinfo{person}{Badrul Sarwar}, \bibinfo{person}{George Karypis}, \bibinfo{person}{Joseph Konstan}, {and} \bibinfo{person}{John Riedl}.} \bibinfo{year}{2001}\natexlab{}.
\newblock \showarticletitle{Item-based collaborative filtering recommendation algorithms}. In \bibinfo{booktitle}{\emph{Proceedings of the 10th international conference on World Wide Web}}. \bibinfo{pages}{285--295}.
\newblock


\bibitem[Sedhain et~al\mbox{.}(2015)]%
        {sedhain2015autorec}
\bibfield{author}{\bibinfo{person}{Suvash Sedhain}, \bibinfo{person}{Aditya~Krishna Menon}, \bibinfo{person}{Scott Sanner}, {and} \bibinfo{person}{Lexing Xie}.} \bibinfo{year}{2015}\natexlab{}.
\newblock \showarticletitle{Autorec: Autoencoders meet collaborative filtering}. In \bibinfo{booktitle}{\emph{Proceedings of the 24th international conference on World Wide Web}}. \bibinfo{pages}{111--112}.
\newblock


\bibitem[Seminario and Wilson(2014)]%
        {seminario2014attacking}
\bibfield{author}{\bibinfo{person}{Carlos~E Seminario} {and} \bibinfo{person}{David~C Wilson}.} \bibinfo{year}{2014}\natexlab{}.
\newblock \showarticletitle{Attacking item-based recommender systems with power items}. In \bibinfo{booktitle}{\emph{Proceedings of the 8th ACM Conference on Recommender systems}}. \bibinfo{pages}{57--64}.
\newblock


\bibitem[Song et~al\mbox{.}(2020)]%
        {song2020poisonrec}
\bibfield{author}{\bibinfo{person}{Junshuai Song}, \bibinfo{person}{Zhao Li}, \bibinfo{person}{Zehong Hu}, \bibinfo{person}{Yucheng Wu}, \bibinfo{person}{Zhenpeng Li}, \bibinfo{person}{Jian Li}, {and} \bibinfo{person}{Jun Gao}.} \bibinfo{year}{2020}\natexlab{}.
\newblock \showarticletitle{Poisonrec: an adaptive data poisoning framework for attacking black-box recommender systems}. In \bibinfo{booktitle}{\emph{2020 IEEE 36th International Conference on Data Engineering (ICDE)}}. IEEE, \bibinfo{pages}{157--168}.
\newblock


\bibitem[Su and Khoshgoftaar(2009)]%
        {su2009survey}
\bibfield{author}{\bibinfo{person}{Xiaoyuan Su} {and} \bibinfo{person}{Taghi~M. Khoshgoftaar}.} \bibinfo{year}{2009}\natexlab{}.
\newblock \showarticletitle{A Survey of Collaborative Filtering Techniques}.
\newblock \bibinfo{journal}{\emph{Advances in artificial intelligence}} (\bibinfo{date}{Jan.} \bibinfo{year}{2009}).
\newblock


\bibitem[Tang et~al\mbox{.}(2020)]%
        {tang2020revisiting}
\bibfield{author}{\bibinfo{person}{Jiaxi Tang}, \bibinfo{person}{Hongyi Wen}, {and} \bibinfo{person}{Ke Wang}.} \bibinfo{year}{2020}\natexlab{}.
\newblock \showarticletitle{Revisiting adversarially learned injection attacks against recommender systems}. In \bibinfo{booktitle}{\emph{Proceedings of the 14th ACM Conference on Recommender Systems}}. \bibinfo{pages}{318--327}.
\newblock


\bibitem[Wang et~al\mbox{.}(2019)]%
        {wang2019neural}
\bibfield{author}{\bibinfo{person}{Xiang Wang}, \bibinfo{person}{Xiangnan He}, \bibinfo{person}{Meng Wang}, \bibinfo{person}{Fuli Feng}, {and} \bibinfo{person}{Tat-Seng Chua}.} \bibinfo{year}{2019}\natexlab{}.
\newblock \showarticletitle{Neural graph collaborative filtering}. In \bibinfo{booktitle}{\emph{Proceedings of the 42nd international ACM SIGIR conference on Research and development in Information Retrieval}}. \bibinfo{pages}{165--174}.
\newblock


\bibitem[Wang et~al\mbox{.}(2023)]%
        {wang2023poisoning}
\bibfield{author}{\bibinfo{person}{Yanling Wang}, \bibinfo{person}{Yuchen Liu}, \bibinfo{person}{Qian Wang}, \bibinfo{person}{Cong Wang}, {and} \bibinfo{person}{Chenliang Li}.} \bibinfo{year}{2023}\natexlab{}.
\newblock \showarticletitle{Poisoning Self-supervised Learning Based Sequential Recommendations}. In \bibinfo{booktitle}{\emph{Proceedings of the 46th International ACM SIGIR Conference on Research and Development in Information Retrieval}}. \bibinfo{pages}{300--310}.
\newblock


\bibitem[Wu et~al\mbox{.}(2021)]%
        {wu2021triple}
\bibfield{author}{\bibinfo{person}{Chenwang Wu}, \bibinfo{person}{Defu Lian}, \bibinfo{person}{Yong Ge}, \bibinfo{person}{Zhihao Zhu}, {and} \bibinfo{person}{Enhong Chen}.} \bibinfo{year}{2021}\natexlab{}.
\newblock \showarticletitle{Triple adversarial learning for influence based poisoning attack in recommender systems}. In \bibinfo{booktitle}{\emph{Proceedings of the 27th ACM SIGKDD Conference on Knowledge Discovery \& Data Mining}}. \bibinfo{pages}{1830--1840}.
\newblock


\bibitem[Wu et~al\mbox{.}(2023a)]%
        {wu2023influence}
\bibfield{author}{\bibinfo{person}{Chenwang Wu}, \bibinfo{person}{Defu Lian}, \bibinfo{person}{Yong Ge}, \bibinfo{person}{Zhihao Zhu}, {and} \bibinfo{person}{Enhong Chen}.} \bibinfo{year}{2023}\natexlab{a}.
\newblock \showarticletitle{Influence-Driven Data Poisoning for Robust Recommender Systems}.
\newblock \bibinfo{journal}{\emph{IEEE Transactions on Pattern Analysis and Machine Intelligence}} (\bibinfo{year}{2023}).
\newblock


\bibitem[Wu et~al\mbox{.}(2023b)]%
        {wu2023attacking}
\bibfield{author}{\bibinfo{person}{Yiqing Wu}, \bibinfo{person}{Ruobing Xie}, \bibinfo{person}{Zhao Zhang}, \bibinfo{person}{Yongchun Zhu}, \bibinfo{person}{FuZhen Zhuang}, \bibinfo{person}{Jie Zhou}, \bibinfo{person}{Yongjun Xu}, {and} \bibinfo{person}{Qing He}.} \bibinfo{year}{2023}\natexlab{b}.
\newblock \showarticletitle{Attacking Pre-trained Recommendation}.
\newblock \bibinfo{journal}{\emph{arXiv preprint arXiv:2305.03995}} (\bibinfo{year}{2023}).
\newblock


\bibitem[Yang et~al\mbox{.}(2017)]%
        {yang2017fake}
\bibfield{author}{\bibinfo{person}{Guolei Yang}, \bibinfo{person}{Neil~Zhenqiang Gong}, {and} \bibinfo{person}{Ying Cai}.} \bibinfo{year}{2017}\natexlab{}.
\newblock \showarticletitle{Fake Co-visitation Injection Attacks to Recommender Systems.}. In \bibinfo{booktitle}{\emph{NDSS}}.
\newblock


\bibitem[Ying et~al\mbox{.}(2018)]%
        {ying2018graph}
\bibfield{author}{\bibinfo{person}{Rex Ying}, \bibinfo{person}{Ruining He}, \bibinfo{person}{Kaifeng Chen}, \bibinfo{person}{Pong Eksombatchai}, \bibinfo{person}{William~L Hamilton}, {and} \bibinfo{person}{Jure Leskovec}.} \bibinfo{year}{2018}\natexlab{}.
\newblock \showarticletitle{Graph convolutional neural networks for web-scale recommender systems}. In \bibinfo{booktitle}{\emph{Proceedings of the 24th ACM SIGKDD International Conference on Knowledge Discovery and Data Mining}} \emph{(\bibinfo{series}{KDD '18})}. \bibinfo{pages}{974--983}.
\newblock


\bibitem[Yuan et~al\mbox{.}(2022)]%
        {yuan2022tenrec}
\bibfield{author}{\bibinfo{person}{Guanghu Yuan}, \bibinfo{person}{Fajie Yuan}, \bibinfo{person}{Yudong Li}, \bibinfo{person}{Beibei Kong}, \bibinfo{person}{Shujie Li}, \bibinfo{person}{Lei Chen}, \bibinfo{person}{Min Yang}, \bibinfo{person}{Chenyun Yu}, \bibinfo{person}{Bo Hu}, \bibinfo{person}{Zang Li}, {et~al\mbox{.}}} \bibinfo{year}{2022}\natexlab{}.
\newblock \showarticletitle{Tenrec: A Large-scale Multipurpose Benchmark Dataset for Recommender Systems}.
\newblock \bibinfo{journal}{\emph{Advances in Neural Information Processing Systems}}  \bibinfo{volume}{35} (\bibinfo{year}{2022}), \bibinfo{pages}{11480--11493}.
\newblock


\bibitem[Yue et~al\mbox{.}(2021)]%
        {yue2021black}
\bibfield{author}{\bibinfo{person}{Zhenrui Yue}, \bibinfo{person}{Zhankui He}, \bibinfo{person}{Huimin Zeng}, {and} \bibinfo{person}{Julian McAuley}.} \bibinfo{year}{2021}\natexlab{}.
\newblock \showarticletitle{Black-box attacks on sequential recommenders via data-free model extraction}. In \bibinfo{booktitle}{\emph{Proceedings of the 15th ACM Conference on Recommender Systems}}. \bibinfo{pages}{44--54}.
\newblock


\bibitem[Zhang et~al\mbox{.}(2020b)]%
        {zhang2020practical}
\bibfield{author}{\bibinfo{person}{Hengtong Zhang}, \bibinfo{person}{Yaliang Li}, \bibinfo{person}{Bolin Ding}, {and} \bibinfo{person}{Jing Gao}.} \bibinfo{year}{2020}\natexlab{b}.
\newblock \showarticletitle{Practical data poisoning attack against next-item recommendation}. In \bibinfo{booktitle}{\emph{Proceedings of The Web Conference 2020}}. \bibinfo{pages}{2458--2464}.
\newblock


\bibitem[Zhang et~al\mbox{.}(2021a)]%
        {zhang2021data}
\bibfield{author}{\bibinfo{person}{Hengtong Zhang}, \bibinfo{person}{Changxin Tian}, \bibinfo{person}{Yaliang Li}, \bibinfo{person}{Lu Su}, \bibinfo{person}{Nan Yang}, \bibinfo{person}{Wayne~Xin Zhao}, {and} \bibinfo{person}{Jing Gao}.} \bibinfo{year}{2021}\natexlab{a}.
\newblock \showarticletitle{Data poisoning attack against recommender system using incomplete and perturbed data}. In \bibinfo{booktitle}{\emph{Proceedings of the 27th ACM SIGKDD Conference on Knowledge Discovery \& Data Mining}}. \bibinfo{pages}{2154--2164}.
\newblock


\bibitem[Zhang et~al\mbox{.}(2023)]%
        {zhang2023robust}
\bibfield{author}{\bibinfo{person}{Kaike Zhang}, \bibinfo{person}{Qi Cao}, \bibinfo{person}{Fei Sun}, \bibinfo{person}{Yunfan Wu}, \bibinfo{person}{Shuchang Tao}, \bibinfo{person}{Huawei Shen}, {and} \bibinfo{person}{Xueqi Cheng}.} \bibinfo{year}{2023}\natexlab{}.
\newblock \showarticletitle{Robust Recommender System: A Survey and Future Directions}.
\newblock \bibinfo{journal}{\emph{arXiv preprint arXiv:2309.02057}} (\bibinfo{year}{2023}).
\newblock


\bibitem[Zhang and Kim(2023)]%
        {zhang2023survey}
\bibfield{author}{\bibinfo{person}{Peiyan Zhang} {and} \bibinfo{person}{Sunghun Kim}.} \bibinfo{year}{2023}\natexlab{}.
\newblock \showarticletitle{A survey on incremental update for neural recommender systems}.
\newblock \bibinfo{journal}{\emph{arXiv preprint arXiv:2303.02851}} (\bibinfo{year}{2023}).
\newblock


\bibitem[Zhang et~al\mbox{.}(2020a)]%
        {zhang2020retrain}
\bibfield{author}{\bibinfo{person}{Yang Zhang}, \bibinfo{person}{Fuli Feng}, \bibinfo{person}{Chenxu Wang}, \bibinfo{person}{Xiangnan He}, \bibinfo{person}{Meng Wang}, \bibinfo{person}{Yan Li}, {and} \bibinfo{person}{Yongdong Zhang}.} \bibinfo{year}{2020}\natexlab{a}.
\newblock \showarticletitle{How to retrain recommender system? A sequential meta-learning method}. In \bibinfo{booktitle}{\emph{Proceedings of the 43rd International ACM SIGIR Conference on Research and Development in Information Retrieval}}. \bibinfo{pages}{1479--1488}.
\newblock


\bibitem[Zhang et~al\mbox{.}(2021b)]%
        {zhang2021reverse}
\bibfield{author}{\bibinfo{person}{Yihe Zhang}, \bibinfo{person}{Xu Yuan}, \bibinfo{person}{Jin Li}, \bibinfo{person}{Jiadong Lou}, \bibinfo{person}{Li Chen}, {and} \bibinfo{person}{Nian-Feng Tzeng}.} \bibinfo{year}{2021}\natexlab{b}.
\newblock \showarticletitle{Reverse Attack: Black-box Attacks on Collaborative Recommendation}. In \bibinfo{booktitle}{\emph{Proceedings of the 2021 ACM SIGSAC Conference on Computer and Communications Security}}. \bibinfo{pages}{51--68}.
\newblock


\end{thebibliography}
\clearpage
\nobalance


\end{document}